%% file: main.tex
\documentclass[10pt, conference]{IEEEtran}
\IEEEoverridecommandlockouts

\usepackage{enumitem}
\usepackage{mathtools}
\usepackage{graphicx}
\usepackage{multirow}
\usepackage{flushend}
\usepackage{xspace}
\usepackage{subcaption}
\usepackage{booktabs}
\usepackage{tcolorbox}
\usepackage{array}
\usepackage{listings}

\usepackage[numbers,sort]{natbib}
\usepackage[colorlinks=true,citecolor=blue,linkcolor=blue,urlcolor=blue,bookmarks=false]{hyperref}

\usepackage{amssymb}
\usepackage{pifont}
\newcommand{\cmark}{\ding{51}}%
\newcommand{\xmark}{\ding{55}}%

\usepackage{microtype}
\newcommand{\tabincell}[2]{\begin{tabular}{@{}#1@{}}#2\end{tabular}}
\newcommand{\todo}[1]{\textcolor{black}{#1}}
\newcommand{\tool}{\textsc{Leopard}\xspace}
\newcommand{\ning}[1]{\textcolor{black}{#1}}
\newcommand{\crc}[1]{\textcolor{black}{#1}}

\newcommand{\fuzzer}{\emph{FOT}\xspace}
\newcommand{\bind}{BIND 9.11.0}
\newcommand{\binutils}{Binutils 2.28}
\newcommand{\ffmpeg}{FFmpeg 3.1.3}
\newcommand{\freetype}{FreeType 2.3.9}
\newcommand{\libav}{Libav 11.8}
\newcommand{\libtiff}{LibTIFF 4.0.6}
\newcommand{\libxslt}{libxslt 1.1.28}
\newcommand{\linux}{Linux 4.12.8}
\newcommand{\openssl}{OpenSSL 1.0.1t}
\newcommand{\sqlite}{SQLite 3.8.2}
\newcommand{\wireshark}{Wireshark 2.2.0}

\definecolor{mGreen}{rgb}{0,0.6,0}
\definecolor{mGray}{rgb}{0.5,0.5,0.5}
\definecolor{mPurple}{rgb}{0.58,0,0.82}

\lstdefinestyle{CStyle}{
	frame=single,
	commentstyle=\color{mGreen},
	keywordstyle=\color{magenta},
	numberstyle=\tiny\color{mGray},
	stringstyle=\color{mPurple},
	basicstyle=\footnotesize,
	showstringspaces=false,
	tabsize=2,
	numbers=left,
	stepnumber=1,
	language=C,
	xleftmargin=.2in,
	xrightmargin=.05in
}

\begin{document}

\title{\tool: Identifying Vulnerable Code for Vulnerability Assessment through Program Metrics}

\author{
\IEEEauthorblockN{Xiaoning Du\IEEEauthorrefmark{1}\IEEEauthorrefmark{5}, Bihuan Chen\IEEEauthorrefmark{2}\IEEEauthorrefmark{5}, Yuekang Li\IEEEauthorrefmark{1}, Jianmin Guo\IEEEauthorrefmark{3}, Yaqin Zhou\IEEEauthorrefmark{1}, Yang Liu\IEEEauthorrefmark{1}\IEEEauthorrefmark{4}, Yu Jiang\IEEEauthorrefmark{3}}
\IEEEauthorblockA{\IEEEauthorrefmark{1}School of Computer Science and Engineering, Nanyang Technological University, Singapore}
\IEEEauthorblockA{\IEEEauthorrefmark{2}School of Computer Science and Shanghai Key Laboratory of Data Science, Fudan University, China}
\IEEEauthorblockA{\IEEEauthorrefmark{3}KLISS, BNRist, School of Software, Tsinghua University, China}
\IEEEauthorblockA{\IEEEauthorrefmark{4}College of Information Science, Zhejiang Sci-Tech University, China}
\IEEEauthorblockA{\IEEEauthorrefmark{5}Co-First Authors}
}


\maketitle

\begin{abstract}
\input{sec00-abstract}
\end{abstract}


\input{sec01-introduction}
\input{sec02-overview}
\input{sec03-methodology}

\input{sec041-application}

\input{sec0411-application-manual}

\input{sec0412-application-fuzzing}
\input{sec0413-application-fuzzing}
\input{sec0402-evaluation}

\input{sec0401-evaluation}

\input{sec0403-evaluation-manual-auditing}

\input{sec0404-evaluation-fuzzing}
\input{sec05-discuss}

\input{sec06-related-work}
\input{sec07-conclusions}
\section{Acknowledgment}
This research was supported (in part)~by~the National Research Foundation, Prime Ministers Office, Singapore under its National Cybersecurity R\&D Program (Award No. NRF2014NCR-NCR001-30, Award No. NRF2016NCR-NCR002-026) and administered by the National Cybersecurity R\&D Directorate, and and Alibaba Group through Alibaba Innovative Research (AIR) Program.

{\footnotesize
\bibliographystyle{IEEEtranS}
\bibliography{IEEEabrv,reference}
}

\end{document}

%% file: sec00-abstract.tex
Identifying potentially vulnerable locations in a code base is critical as a pre-step for effective vulnerability assessment; i.e.,~it~can~greatly help security experts put their time~and~effort~to where~it~is~needed most. Metric-based~and~pattern-based methods have been presented for identifying vulnerable code. The former relies on machine learning and cannot work well due~to~the~severe imbalance between non-vulnerable and vulnerable code~or~lack~of features to characterize vulnerabilities. The~latter~needs~the prior knowledge~of~known vulnerabilities and can only identify similar but not new types of vulnerabilities.


In this paper, we propose and implement a generic,~lightweight and extensible framework, \tool, to identify potentially vulnerable functions through program metrics. \tool~requires no prior knowledge about known vulnerabilities.~It~has two steps by combining two sets of systematically derived metrics.
First, it uses complexity metrics to group the functions~in~a~target~application~into~a~set~of~bins. Then, it uses~vulnerability~metrics~to rank the functions in each bin and identifies~the~top~ones~as~potentially vulnerable.
Our experimental results on 11 real-world~projects have demonstrated that, \tool can cover \ning{74.0\%} of \ning{vulnerable functions} by identifying \ning{20\%} of functions as vulnerable~and~outperform machine learning-based and static analysis-based techniques. 
We further propose three applications of \tool for manual code review and fuzzing, through which we discovered 22 new bugs in real applications like~{\tt PHP}, {\tt radare2} and {\tt FFmpeg}, and eight of them are new vulnerabilities. 

%% file: sec01-introduction.tex

\section{Introduction}\label{sec:intro}

Vulnerabilities are one of the key threats to software~security \cite{Meng2015CSS}. Security experts usually leverage guided fuzzing~(e.g.,~\cite{Wang2010, Woo2013, Bohme2016, Rawat2017}),~symbolic execution (e.g., \cite{GodefroidL2008, Babic2011, Cha2015, Stephens2016})~or~manual auditing to hunt vulnerabilities. As only~a~few vulnerabilities are scattered~across~a~large code base, vulnerability~hunting is a very challenging task~that requires intensive knowledge and is comparable to finding ``a needle in a haystack''~\cite{Zimmermann2010}. Therefore, a large amount~of~time and effort is wasted in analyzing the non-vulnerable code. In that sense, identifying potentially vulnerable code in a code base can guide vulnerability hunting and assessment in a promising direction.

There are two types of existing techniques to automatically identify vulnerabilities: metric-based and pattern-based techniques. Metric-based techniques, inspired by bug~prediction \cite{Gyimothy2005, Catal2009, Hall2012, Radjenovic2013, Malhotra2015, nam2015clami, zhang2016cross}, leverage supervised or unsupervised machine learning to predict vulnerable code mostly~at~the~granularity level of a source file. Following security experts' belief that complexity is the enemy of software~security~\cite{McGraw2006}, they use complexity metrics~\cite{Shin2008a, Shin2008b, Chowdhury2011, Moshtari2013, Moshtari2016} as features, or combine them with code churn metrics~\cite{Gegick2008, Shin2011a,Shin2013}, token frequency metrics~\cite{Scandariato2014, Walden2014, Zhang2015, Hovsepyan2016}, dependency metrics~\cite{Neuhaus2007, Nguyen2010, Zimmermann2010, Morrison2015}, developer activity metrics \cite{Shin2011a, Shin2013} and execution complexity metrics~\cite{Shin2011b}. On the other hand, pattern-based techniques leverage patterns of known vulnerabilities to identify potentially vulnerable code through static analysis. The patterns~are~formulated based on the syntax or semantics abstraction of a certain type of vulnerabilities, e.g., missing security checks on security-critical objects~\cite{Son2011, yamaguchi2013chucky}, security properties~\cite{Vanegue2013}, code structures~\cite{Yamaguchi2012}, and vulnerability specifications \cite{Livshits2005,Yamaguchi2014}. 

While vulnerability identification has been attracting~great attention, some problems still remain. On one hand, metric-based~techniques are mostly designed for one single application (or a few~applications of the same type). Thus, they might~not work on a variety~of diverse applications as machine learning may over-fit to noise features. Moreover, while an empirical connection between vulnerabilities and bugs exist, the connection is considerably weak due to the differences between vulnerabilities and bugs \cite{Camilo2015}. As a result,~the~research on bug prediction  cannot directly translate~to~vulnerability identification. Unfortunately, the existing metric-based techniques use the similar metrics~as~those in bug prediction, and thus fail~to~investigate the characteristics of vulnerabilities.



On the other hand, metric-based and pattern-based techniques mostly require a great deal of prior knowledge about vulnerabilities. In~particular, a large number of known vulnerabilities are needed for~effective supervised machine learning in some metric-based techniques. The number of vulnerabilities~is~much smaller than the number of bugs, and the imbalance between non-vulnerable and vulnerable code is severe, which hinders the applicability~of~supervised machine learning to vulnerable code identification. Similarly, a prerequisite of those pattern-based techniques is the existence of known vulnerabilities as the guideline to formulate patterns. They can~only identify~similar but not new vulnerabilities. Further, patterns are often application-specific, and~thus those techniques are usually used as in-project but not cross-project vulnerable code identification.

In this paper, we propose a vulnerability identification framework, named \tool\footnote{Leopard is known for its opportunistic hunting behavior, broad diet, and strength, which reflect the identification capabilities we are pursuing.}, to identify potentially vulnerable functions in C/C++ applications. \tool is designed to be {\it generic}~to~work~for different types~of~applications, {\it lightweight} to support the analysis of large-scale applications and {\it extensible} with domain-specific~data~to improve~the~accuracy. We design \tool as a pre-step for vulnerability assessment, but not to directly pinpoint vulnerabilities. We propose three different applications of \tool to guide security~experts during the manual auditing or automatic fuzzing by narrowing down the space of potentially vulnerable functions.



 \tool does not require any prior knowledge about known vulnerabilities. It works in two steps by combining two sets~of systematically derived program metrics, i.e., complexity metrics and vulnerability metrics. Complexity metrics capture~the~complexity of a function in two complementary dimensions:~the cyclomatic complexity of the function, and~the~loop structures in the function.
Vulnerability metrics~reflect~the~vulnerable~characteristics of functions in three dimensions: the dependency of the function, pointer usage in the function, and the dependency among control structures within the function.


\tool first uses complexity metrics to group the functions in a target application into a set of bins. Then, \tool leverages vulnerability metrics to rank the functions in each bin and identify the top functions~in~each bin as potentially vulnerable. We propose such a binning-and-ranking approach as there often exists a \crc{proportional relation} between complexity and vulnerability metrics, which is evidenced~in~our~experimental study. As a result, each bin has a different level of complexity, and our framework can identify vulnerabilities at all levels of complexity without missing low-complexity ones.

We implemented the proposed framework to obtain complexity and vulnerability metrics for C/C++ programs. 
We evaluated the effectiveness and scalability of our framework with \ning{11} different types of real-world projects. \tool can cover \ning{74.0\%} of \ning{vulnerable~functions} by identifying \ning{20\%} of functions as potentially vulnerable, outperforming both typical machine learning-based and static analysis-based techniques. Applying \tool on~{\tt PHP}, {\tt MJS}, {\tt XED}, {\tt FFmpeg} and {\tt Radare2} and with further manual auditing or automatic fuzzing, we discovered 22 new bugs, among which eight are new vulnerabilities.

In summary, our work makes the following contributions.
\begin{itemize}[leftmargin=*]
\item We propose a generic, lightweight and extensible framework to identify potentially vulnerable functions, which incorporates two~sets of program metrics.
\item We propose three different applications of \tool to guide security~experts during the manual auditing or automatic fuzzing to hunt for vulnerabilities.
\item We implemented our framework and conducted large-scale experiments on 11 real-world projects~to~demonstrate~the~effectiveness and scalability of our framework.
\item We demonstrated three application scenarios of our framework and found 22 new bugs.
\end{itemize}


%% file: sec02-overview.tex

\section{Methodology}\label{sec:methodology}

In this section, we present the overview of \tool and elaborate each step of the proposed approach.

\subsection{Overview}\label{sec:overview}

\begin{figure}[!t]
\centering
\includegraphics[width=0.45\textwidth]{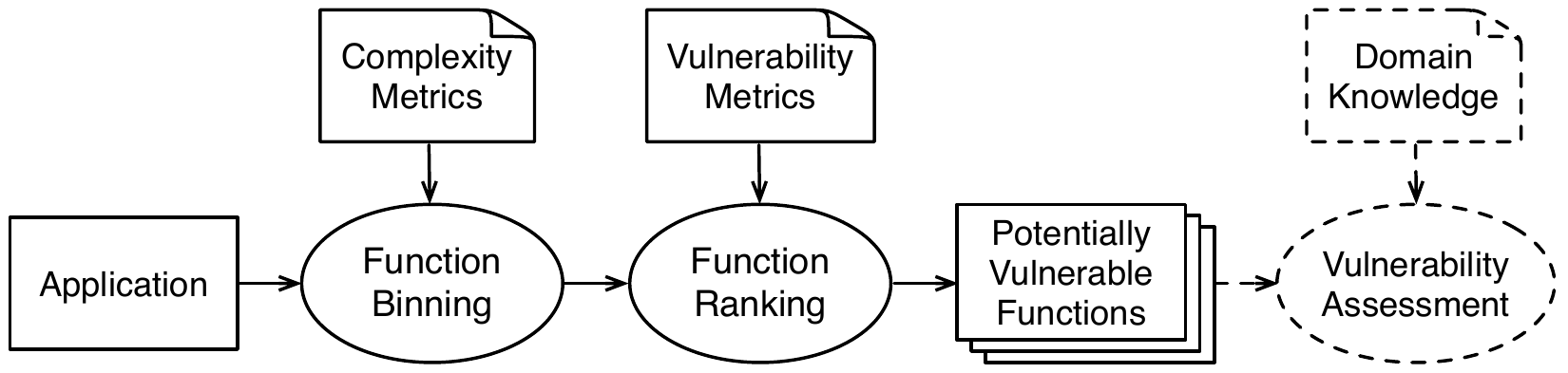}
\vspace{-3pt}
\caption{An Overview of the Proposed Framework}
\label{fig:overview}
\end{figure}

Fig.~\ref{fig:overview} presents the work flow of \tool, which is designed to be generic, lightweight and extensible. The input is the source code~of~a C/C++ application. \tool works~in~two~steps: function binning and function ranking, and returns~a~list~of~potentially vulnerable functions for vulnerability assessment. 

In the first step (\S~\ref{sec:binning}), we use complexity metrics to group all functions 
in the target application into a set of bins. The complexity metrics capture the complexity of a function in two dimensions: the function itself (i.e., cyclomatic complexity) and the loop structures in the function (e.g., the number of nested loops).
Each bin has a different level of complexity, which is designed to identify vulnerabilities~at all levels of complexity (i.e., avoid missing vulnerable functions with low-complexity).

In the second step (\S~\ref{sec:ranking}), we use vulnerability~metrics to rank the functions~in each bin in order to identify~the~top functions~in~each bin as potentially vulnerable. The vulnerability~metrics~capture~the vulnerable characteristics of a function in three dimensions: the dependency of the function (e.g.,~the number~of parameters), the pointer usage in a function (e.g., the number of pointer arithmetic) and the dependency~of~control structures in the function (e.g., the number of nested control structures).
By incorporating such metrics, we can have a high potential of characterizing and identifying vulnerable functions.

\tool is designed to support and facilitate confirmative vulnerability assessments,~e.g., to guide security experts during automatic~fuzzing \cite{Wang2010,Woo2013,Bohme2016,Rawat2017}~or~manual auditing by providing potentially vulnerable function list and the corresponding metrics information. With such~knowledge, security experts~can~prioritize the assessment order, choose~the appropriate analysis technique, and analyze the root cause. Further, based on application-specific~domain knowledge (e.g., vulnerability history and heavily fuzzed function lists), security experts can further rank or filter the potentially vulnerable functions~to~focus on those more interesting functions.

Using program metrics in a simple binning-and-ranking way makes \tool satisfy our design principle of being~generic and lightweight. It is applicable to any large-scale applications of any type and does not require~prior knowledge about known vulnerabilities.
The two sets of metrics are comprehensive, but also are extensible with new metrics as we gather more~usage feedback from security experts (see discussion in \S~\ref{sec:discuss}). Thus, \tool also satisfies our design principle of being extensible such that it can be further enhanced.


%% file: sec03-methodology.tex



\subsection{Function Binning}\label{sec:binning}

Different vulnerabilities often have different levels of complexity.
To identify vulnerabilities at all levels of complexity, in the~first~step, we categorize all functions
in~the~target application into a set~of~bins based on complexity metrics.
As a result,~each bin represents~a~different level of complexity.
Afterwards, the second step (\S~\ref{sec:ranking}) plays the prediction role via ranking. Such a binning-and-ranking approach is designed to avoid missing low-complexity vulnerable functions.

\noindent \textit{\textbf{Complexity Metrics.}} 
By ``complexity", we refer to the approximate number of paths in a function, and derive~the~complexity metrics of a function from its structural complexity.
A function often has loop and control structures, which are the main sources of structural complexity.
Cyclomatic complexity \cite{McCabe1976} is a widely-used metric to measure the complexity,~but without reflection of the loop structures.
Based~on such understanding, we introduce the complexity of a function~with~respect to these two complementary dimensions, as shown in Table~\ref{table:complex}.

 \textbf{Function metric} (C1) captures the standard cyclomatic complexity~\cite{McCabe1976} of a function, i.e., the number of linearly independent paths through a function. A higher value of C1 means that the function is likely more~difficult~to~analyze~or~test.

 \textbf{Loop structure metrics} (C2--C4) reflect the complexity resulting from loops, which can drastically increase the number of paths in the function.
Metrics include the number~of loops, the number of nested loops, and the maximum nesting level of loops. Loops are challenging in program~analysis~\cite{Xie2016}~and hinder vulnerability analysis.
Basically, the higher these metrics, the more (and possibly longer) paths need to be considered and the more difficult to analyze the function.


\begin{table}[!t]
\centering
\scriptsize
\caption{Complexity Metrics of a Function}\label{table:complex}
\vspace{-5pt}
\begin{tabular}{|c|c|l|}
\hline
Dimension & ID & Metric Description \\\hline\hline
\tabincell{c}{CD1: Function} & C1 & Cyclomatic complexity \\\hline\hline
\multirow{3}*{\tabincell{c}{CD2: Loop Structures}} & C2 & \# of loops \\\cline{2-3}
& C3 & \# of nested loops \\\cline{2-3}
& C4 & Maximum nesting level of loops \\\hline
\end{tabular}
\end{table}

\noindent \textit{\textbf{Binning Strategy.}} Given the values of these complexity metrics for functions in the target application, we compute a {\it complexity score} for each function by adding up all the complexity metric values, and then group~the functions with the same score into the same bin. Here we do not use a range-based binning strategy (i.e., grouping the functions whose scores fall into the same range into the same bin) as it is hard to determine the suitable granularity of the range. Such a simple strategy not only makes our framework lightweight, but also works well, as evidenced by our experimental study in \S~\ref{sec:EffBR}.


\noindent \subsection{Function Ranking}\label{sec:ranking}


Different from the structural complexity metrics,~in~the second step, we derive a new set of vulnerability metrics according~to the characteristics of general causes of vulnerabilities, and then rank the functions and identify the top ones in each bin as potentially vulnerable based on the vulnerability metrics.
\crc{Existing metric-based techniques~\cite{Moshtari2013,Moshtari2016} rarely employ any vulnerability-oriented~metrics, and make no differentiation between complexity metrics and vulnerability metrics.}
Here, we propose and~incorporate vulnerability metrics to have a high potential of characterizing and identifying vulnerable functions.

\noindent \textit{\textbf{Vulnerability Metrics.}} 
Most critical types of vulnerabilities in C/C++ programs are directly or indirectly caused by memory management errors~\cite{Szekeres2013} and/or missing checks on some sensitive variables~\cite{yamaguchi2013chucky} (e.g., pointers).
Resulting vulnerabilities include but are not limited to memory errors, access control errors (e.g., missing checks on user permission), and information leakage.
Actually, the root causes of many denial of service and code execution vulnerabilities can also be traced back to these causes.
The above mentioned types account for more than 70\% of all vulnerabilities~\cite{cvedetails}.
Hence, it is possible to define a set of vulnerability metrics that are compatible with major vulnerability types.
Here we would not favor any specific types of vulnerabilities, e.g., to include metrics like division operation which is closely related to divide-by-zero, while the exploration of type-specific metrics is worth of investigation in the future. 
\crc{With either high or low complexity scores, vulnerable functions we focus on are mainly with complicated and compact computations, which are independent from the number of paths in the function.}
Based~on~these observations, we introduce the vulnerability metrics of a function w.r.t. three dimensions, as summarized in Table~\ref{table:vulnerable}.



\begin{table}[!t]
\centering
\scriptsize
\caption{Vulnerability Metrics of a Function}\label{table:vulnerable}
\vspace{-5pt}
\begin{tabular}{|c|c|l|}
\hline
Dimension & ID & Metric Description \\\hline\hline
\multirow{2}*{\tabincell{c}{VD1: \\Dependency}} & V1 & \# of parameter variables \\\cline{2-3}
& V2 & \# of variables as parameters for callee function \\\hline\hline
\multirow{4}*{\tabincell{c}{VD2: \\Pointers}}
& V3 & \# of pointer arithmetic \\\cline{2-3}
& V4 & \# of variables involved in pointer arithmetic \\\cline{2-3}
& V5 & Max pointer arithmetic a variable is involved in  \\\hline\hline
\multirow{5}*{\tabincell{c}{VD3: \\Control \\Structures}}
& V6 & \# of nested control structures \\\cline{2-3}
& V7 & Maximum nesting level of control structures \\\cline{2-3}
& V8 & Maximum of control-dependent control structures \\\cline{2-3}
& V9 & Maximum of data-dependent control structures \\\cline{2-3}
& V10 & \# of if structures without else \\\cline{2-3}
& V11 & \# of variables involved in control predicates \\\hline
\end{tabular}
\end{table}



 \textbf{Dependency metrics} (V1--V2) characterize the dependency relationship of a function with other functions, i.e., the number of parameter variables of the function and the number of variables prepared by the function as parameters of function calls.
The more dependent with other functions, the more difficult to track the interaction.

 \textbf{Pointer metrics} (V3--V5) capture the manipulation of pointers, i.e.,
~the~number of pointer arithmetic, the number of variables used~in~pointer arithmetic, and the maximum number of pointer~arithmetic a variable is involved in.
Member access operations (e.g., ptr$\rightarrow$m), deference operations (e.g., *ptr), incrementing pointers (e.g., ptr++), and decrementing pointers (e.g., prt-{}-) are all pointer arithmetics.
The number of pointer arithmetic can be obtained from the Abstract Syntax Tree (AST) of the function via simple counting.
These operations are closely related to sensitive memory manipulations, which can increase the risk of memory management errors.

Alongside, we count how many unique variables are used in the pointer arithmetic operations.
The more variables get involved, the more challenging for programmers to make correct decisions.
For these variables, we also examine how many pointer arithmetic operations they are involved in and record the maximum value.
Frequent operations on the same pointer make it harder to track its value and guarantee the correctness.
In a word, the higher these metrics,~the~higher chance to cause complicated memory management problems, and thus higher chance to dereference null or out-of-bound pointers.

 \textbf{Control structure metrics} (V6--V11) capture the vulnerability due to highly coupled and dependent control structures (such as {\it if} and {\it while}), i.e., the number of nested control structures pairs, the maximum nesting level of control structures, the maximum number of control structures that are control- or data-dependent, the number of {\it if} structures without explicit {\it else} statement, and the number of variables that are involved in the data-dependent control structures.
We explain the above metrics with an example (Fig.~\ref{list:fib}) calculating Fibonacci series.
There are two pairs of nested control structures, {\it if} at Line~\ref{line:if2} respectively with {\it if} at Line~\ref{line:if3} and {\it for} at Line~\ref{line:for1}.
Obviously, the maximum nesting level is two, with the outer structure as {\it if} at Line~\ref{line:if2}.
The maximum of control-dependent control structures is 3, including {\it if} at Line~\ref{line:if2} and Line~\ref{line:if3}, and {\it for} at Line~\ref{line:for1}.
The maximum of data-dependent control structures is four since conditions in all four control structures make checks on variable $n$.
All three {\it if} statements are without {\it else}.
There are two variables, i.e., $n$ and $i$ involved in the predicates of control structures.
Actually, the more variables used in the predicates, the more likely to makes error on sanity checks.
The higher these metrics, the harder for programmers to follow, and the more difficult to reach the deeper part of the function during vulnerability hunting. Stand-alone {\it if} structures are suspicious for missing checks on the implicit {\it else} branches.

\begin{figure}[!t]
	\begin{lstlisting}[style=CStyle, basicstyle=\tiny, escapechar=^]
void fibonacci(int *res, int n) {
	if (n <= 0) {^\label{line:if1}^
		return;
	}
	res[0] = 0;
	res[1] = 1;
	if (n > 1) {^\label{line:if2}^
		if (n == 3) {^\label{line:if3}^
			res[2] = 1;
			return;
		}
		for(int i = 2; i <= n; i++) {^\label{line:for1}^
			res[i] = res[i-1] + res[i-2];
		}
	}
}
	\end{lstlisting}
\vspace{-5pt}
\caption{A Function to Calculate Fibonacci Series}
\label{list:fib}
\end{figure}

There usually exists a \crc{proportional relation} between complexity and vulnerability metrics, 
because the more complex the (independent path and loop) structures of a function, the higher chance the variables, pointers and coupled control~structures are involved.
The complexity metrics are used to approximate the number of paths in the function, which are neutral to the vulnerable characteristics.
Importantly, for the set of control structure metrics used as vulnerability indicators, they describe a different aspect of properties than complexity metrics.
First, whether control structures are nested or dependent, or whether {\it if} are followed by {\it else}, are independent to cyclomatic complexity metrics.
Second, intensively coupled control structures are good evidence of vulnerability.
Instead of directly ranking functions with complexity and/or vulnerability metrics, we propose a binning-and-ranking approach to avoid missing less complicated but vulnerable functions, as will be evidenced~in~\S~\ref{sec:RatBR}.

\noindent \textit{\textbf{Ranking Strategy.}} Based on the values of these metrics for~the functions, we compute a {\it vulnerability score} for each~function by adding up all the metric values, rank the functions in each bin according to the scores, and cumulatively identify the top functions with highest scores in each bin as potential vulnerable functions.
During the selection, we identify the top $k$ functions from each bin where $k$ is initially 1,  and increase by 1 in each selection iteration.
Notice that we may take more than $k$ functions as we treat functions with the same score equally.
This selection stops when an appropriate portion (i.e., $p$) of functions has been selected. Here $p$ can be set by users.
Similar~to~the~binning strategy, we adopt a simple ranking strategy to make our framework both lightweight and effective.

%% file: sec041-application.tex
\section{Applications of \tool}\label{sec:application}

\crc{\tool is not designed to directly pinpoint vulnerabilities but to assist confirmative vulnerability assessment.}
\tool outputs a list of potential vulnerable functions with complexity metrics and vulnerability metrics scores, which can provide useful insight for further vulnerability hunting.
In this section, we demonstrate three different ways to apply the results generated by \tool for finding vulnerabilities.
\crc{With \tool, we found 22 new bugs in five widely-used real-world programs.
The detailed experimental results will be introduced in \S~\ref{sec:evalapp}.}

%% file: sec0411-application-manual.tex
\noindent\textit{\textbf {Manual Auditing.}}
In general, with the help of \tool, manual auditing (e.g., code review) can be greatly improved w.r.t. effectiveness and efficiency.
Instead of auditing all~the functions~\cite{Coldwind2017}, security experts can focus on only those potentially vulnerable functions that are identified by \tool.


Furthermore, the vulnerability metrics produced by \tool may help~security experts to quickly identify the root cause of vulnerabilities with their domain knowledge, especially for complicated large functions.
For~example, if a vulnerable function has a large number of instances of \emph{if-without-else}, security experts could pay attention to the logic of the missing \emph{else} to see if there are potential missing checks; and if a vulnerable function has a large number of pointers, security experts could focus on the memory allocation and~deallocation operations to see if there are potential dangling pointers. Although these metrics cannot directly pinpoint the root cause, it can provide explicit hints on the possible root cause.

%

%% file: sec0412-application-fuzzing.tex

\noindent \textit{\textbf {Target Identification for Directed Fuzzing.}}
Fuzzing has been shown as an effective testing technique to find vulnerabilities.
Specifically, greybox fuzzers (e.g., AFL~\cite{afl} and its variants \cite{BohmeNMA17, Bohme2016}) have gained the popularity and been proven to be practical for finding vulnerabilities in real-world applications.


Current greybox fuzzers aim to cover as many program states as possible within a given time budget.
However, higher coverage does not necessarily imply finding more vulnerabilities because fuzzers are blindly exploring all possible program states without focusing the efforts on the more vulnerable functions.
Recently, directed greybox fuzzers (e.g., AFLGo~\cite{BohmeNMA17} and Hawkeye~\cite{Chen:2018:HTD:3243734.3243849}) are proposed to guide the fuzzing execution towards a predefined vulnerable function (a.k.a. target site) to either reproduce the vulnerability or check whether a patched function is still vulnerable~\cite{BohmeNMA17}.

Since \tool produces a list potential vulnerable functions, a straightforward application with directed greybox fuzzers is to set potential vulnerable functions as target sites.
In this way, we can quickly confirm whether a potentially vulnerable function is really vulnerable or a false positive by directing the fuzzer to concentrate on the function. Note that although the fuzzer can reach a vulnerable function, the vulnerability hidden in the function may not always be triggered. But still, directed fuzzing has been shown as an effective technique to reproduce vulnerabilities~\cite{BohmeNMA17}.
\crc{To demonstrate~the idea, we utilize a directed fuzzing tool, Hawkeye~\cite{Chen:2018:HTD:3243734.3243849}, which is built upon an extensible fuzzing framework FOT~\cite{Chen:2018:FVC:3236024.3264593} and reported to outperform ALFGo~\cite{BohmeNMA17}.}
However, due to the large number of the potential vulnerable functions generated by \tool, it is ineffective to set all potential vulnerable functions as target sites as it may confuse the fuzzer where to guide.
To this end, we choose to separate the target application into smaller modules based on its architecture design or simply namespace, and then let the Hawkeye to fuzz with the targets grouped by modules separately.

%% file: sec0413-application-fuzzing.tex
\noindent \textit{\textbf {Seed Prioritization for Fuzzing.}}
Greybox fuzzers often keep interesting test inputs (i.e., seeds) for further fuzzing.
These seeds need to be continuously evaluated to decide which of them should be prioritized.
By default, most fuzzers (e.g., AFL) prefer seeds with ``smaller file size'' and ``shorter execution time'' or ``more edge (basic-block transition) coverage'', which are not vulnerability-aware decisions.

Since \tool assigns each function a vulnerability score and a complexity score, we can use these scores to help to evaluate which seed should be prioritized such that the fuzzer can find more vulnerabilities in the given time budget.
For this purpose, we extended {\fuzzer} by enabling it to accept external function-level scores for seed prioritization.
The detailed seed evaluation process is explained as follows.
First, we calculate a priority score for each function based on the binning-and-ranking strategy.
For a function $\mathcal{F}$ within top $k$, its priority score is calculated using the following formula:
\begin{equation}
\begin{array}{ll@{}r@{}r@{}l}
priority\_score(\mathcal{F}) =  100 - \cfrac{\sum_{i=1}^{k}\mathcal{N}_i}{\mathcal{N}} \cdot 100
\end{array}
\end{equation}
where $\mathcal{N}_i$ is the number of functions with rank $i$ and $\mathcal{N}$ is the total number of all functions.
For example, if the top 1 functions contribute a portion of 20\% to the total number of all functions, then these functions are assigned with a score of 80 ($100 - 20$).
Then, the function-score mapping is provided to \fuzzer.
After executing a test input (i.e., seed), the fuzzer can get an execution trace consist of functions.
Then the fuzzer will accumulate the priority scores of the functions on the execution trace to form the priority score of that trace.
As a result, each seed is associated with a trace priority score representing its \emph{vulnerableness}.
When the fuzzer chooses the next seed to fuzz, it will select the one with highest trace priority score.



%% file: sec0402-evaluation.tex

\begin{table}[t]
\scriptsize
\centering
\caption{Details of the Target Applications}\label{table:project}
\vspace{-5pt}
\begin{tabular}{|c||c|c|c|c|c|}
\hline
Project & \tabincell{c}{SLOC} & \tabincell{c}{\#Func.} & \tabincell{c}{Vul. \\Func.} & \tabincell{c}{CVE} & \tabincell{c}{Excl. \\CVE } \\\hline\hline
\bind & 504K & 9,462 & 9 & 3 & 3\\\hline
\binutils & 3,336K & 24,713 & 84 & 37 & 24\\\hline
\ffmpeg & 986K & 19,336 & 38 & 26 & 6\\\hline
\freetype & 126K & 1,847 & 74 & 48 & 18 \\\hline
\libav & 583k & 10,583 & 8 & 6 & 8 \\\hline
\libtiff & 118K & 1,394 & 20 & 12 & 24 \\\hline
\libxslt & 47K & 666 & 5 & 3 & 1\\\hline
\linux & 17,103K & 488,960 & 256 & 104 & 32 \\\hline
\openssl & 360K & 6,649 & 42 & 17 & 3 \\\hline
\sqlite & 172K & 3,651 & 10 & 7 & 2 \\\hline
\wireshark & 3,551K & 33,564 & 152 & 74 & 31 \\\hline
Total & 26,886K & 600,825 & 698 & 337 & 152\\\hline
\end{tabular}
\end{table}

\section{Evaluation}\label{sec:evaluation}

\tool is implemented in 11K lines of Python code. Specifically, we used Joern \cite{Yamaguchi2014} to extract the values of complexity and vulnerability metrics, given the source code of an application. 
More details of the implementation and evaluation are available at our website \cite{Leopard}.

\subsection{Evaluation Setup}


\noindent \textit{\textbf{Target Applications.}} We used \ning{11} real-world open-source projects that represent a diverse set of applications.
{\tt BIND} is the most widely used Domain Name System (DNS) software.
{\tt Binutils} is a collection of binary tools.
{\tt FFmpeg} is the leading multimedia framework.
{\tt FreeType} is a library to render fonts.
{\tt Libav} is a library for handling multimedia data, which was originally forked from {\tt FFmpeg}.
{\tt LibTIFF} is a library for reading and writing Tagged Image File Format (TIFF) files.
{\tt libxslt} is the XSLT C library for the GNOME project.
{\tt Linux} is a monolithic Unix-like computer operating system kernel.
{\tt OpenSSL} is a robust and full-featured toolkit for the Transport Layer Security (TLS) and Secure Sockets Layer (SSL) protocols.
{\tt SQLite} is a relational database management system.
{\tt Wireshark} is a network traffic analyzer for Unix and Unix-like operating systems.

The details of each target application are reported in Table \ref{table:project}. The first column gives the project version, the second column reports the source lines of code, and the third column lists the total number of functions in each project. The last three columns report the number of vulnerable functions, CVEs
(Common Vulnerabilities and Exposures), and CVEs excluded from our research, collected as ground truth (see below). Here, we chose the recent versions of the projects that had large number of CVEs. The number of functions ranges from \ning{666} for {\tt libxslt} to \ning{488,960} for {\tt Linux}, which is diverse enough to show the generality of our framework.
In total, 26,886K lines of code and 600,825 functions are studied, which makes our study large-scale and its results reliable.


\begin{figure*}[!t]
\centering

\begin{subfigure}[b]{0.19\textwidth}
\centering
\begin{minipage}{\linewidth}
\includegraphics[width=0.9\textwidth]{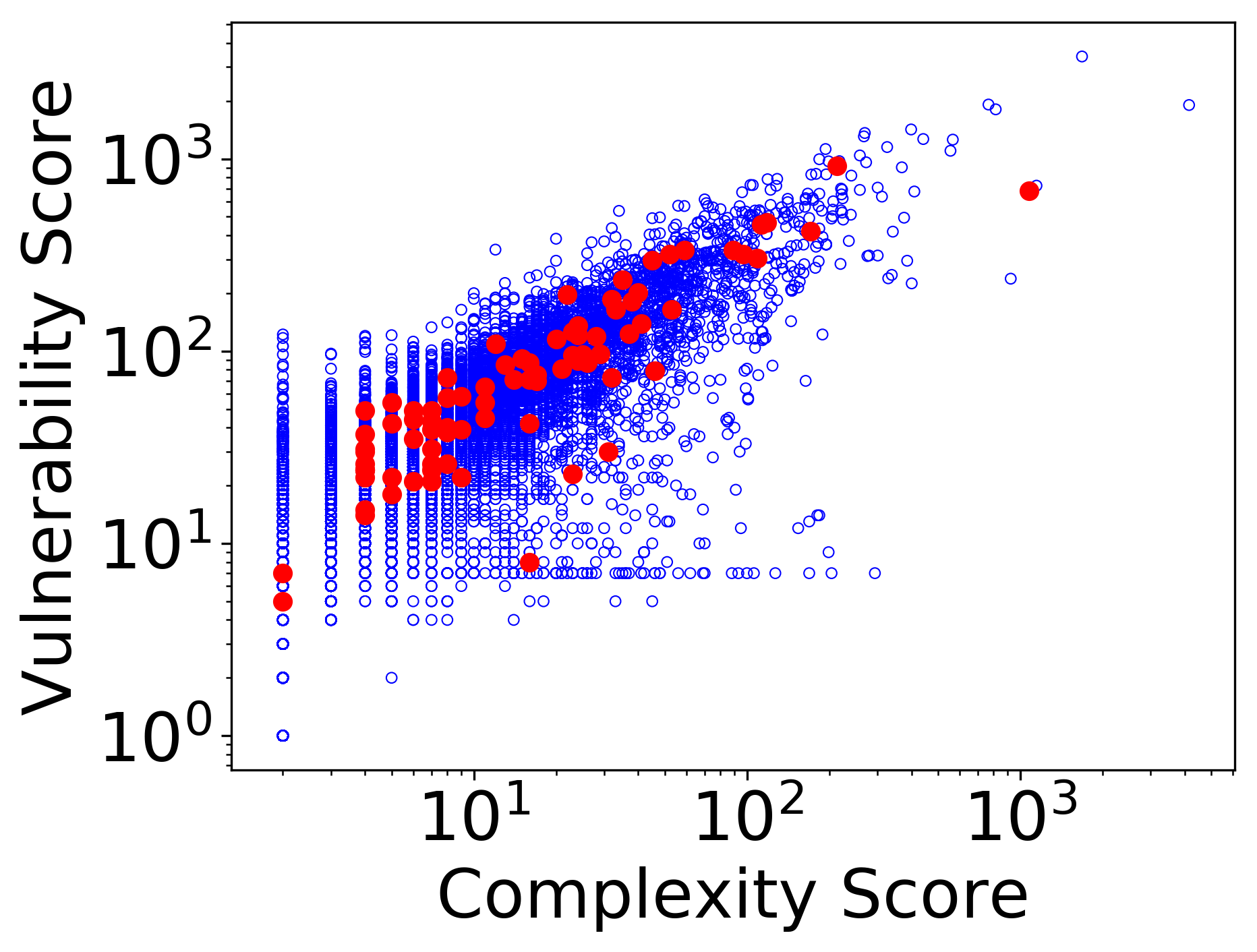}
\end{minipage}
\caption{\binutils}\label{fig:binutils}
\end{subfigure}
\begin{subfigure}[b]{0.19\textwidth}
\centering
\begin{minipage}{\linewidth}
\includegraphics[width=0.9\textwidth]{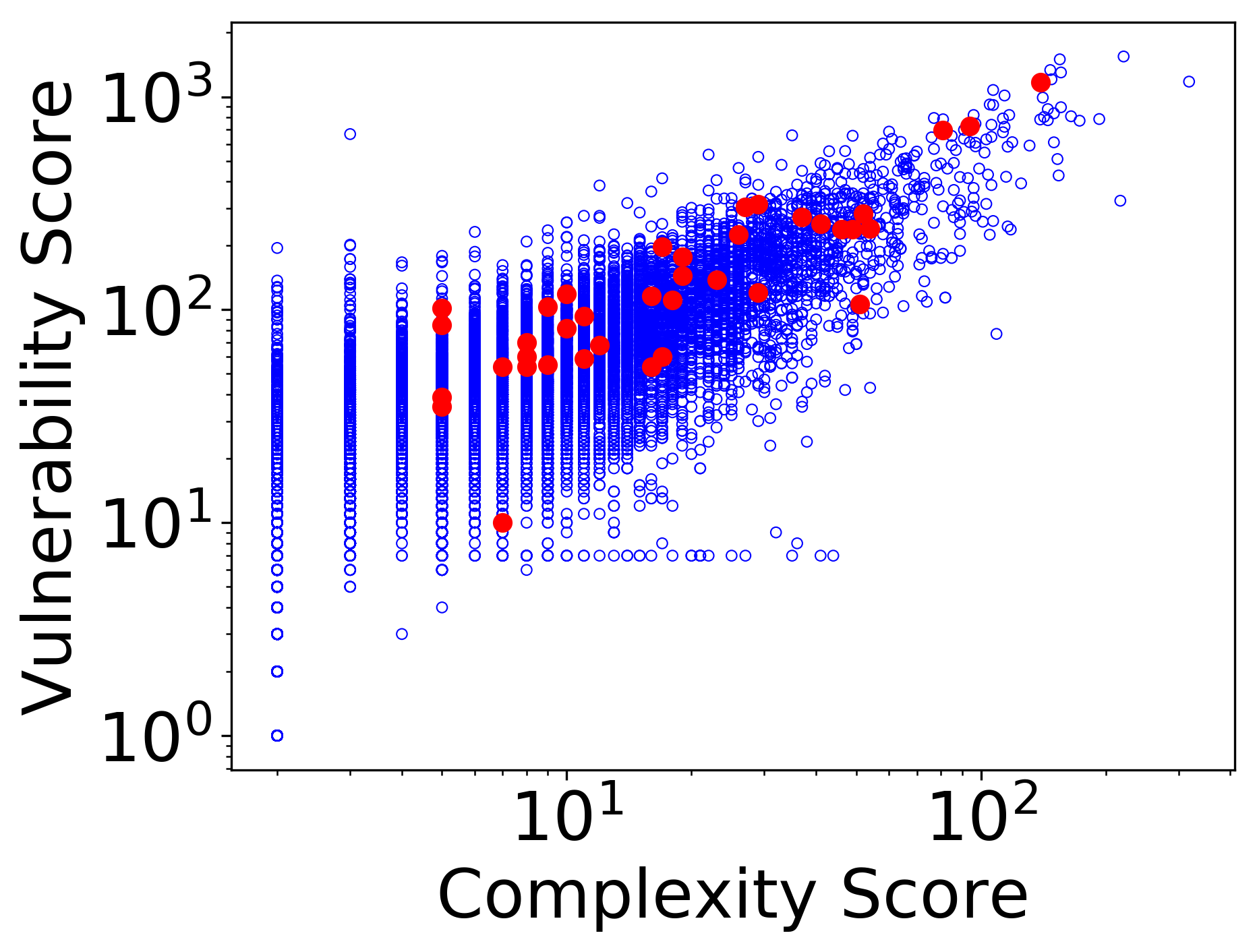}
\end{minipage}
\caption{\ffmpeg}\label{fig:ffmpeg}
\end{subfigure}
\begin{subfigure}[b]{0.19\textwidth}
\centering
\begin{minipage}{\linewidth}
\includegraphics[width=0.9\textwidth]{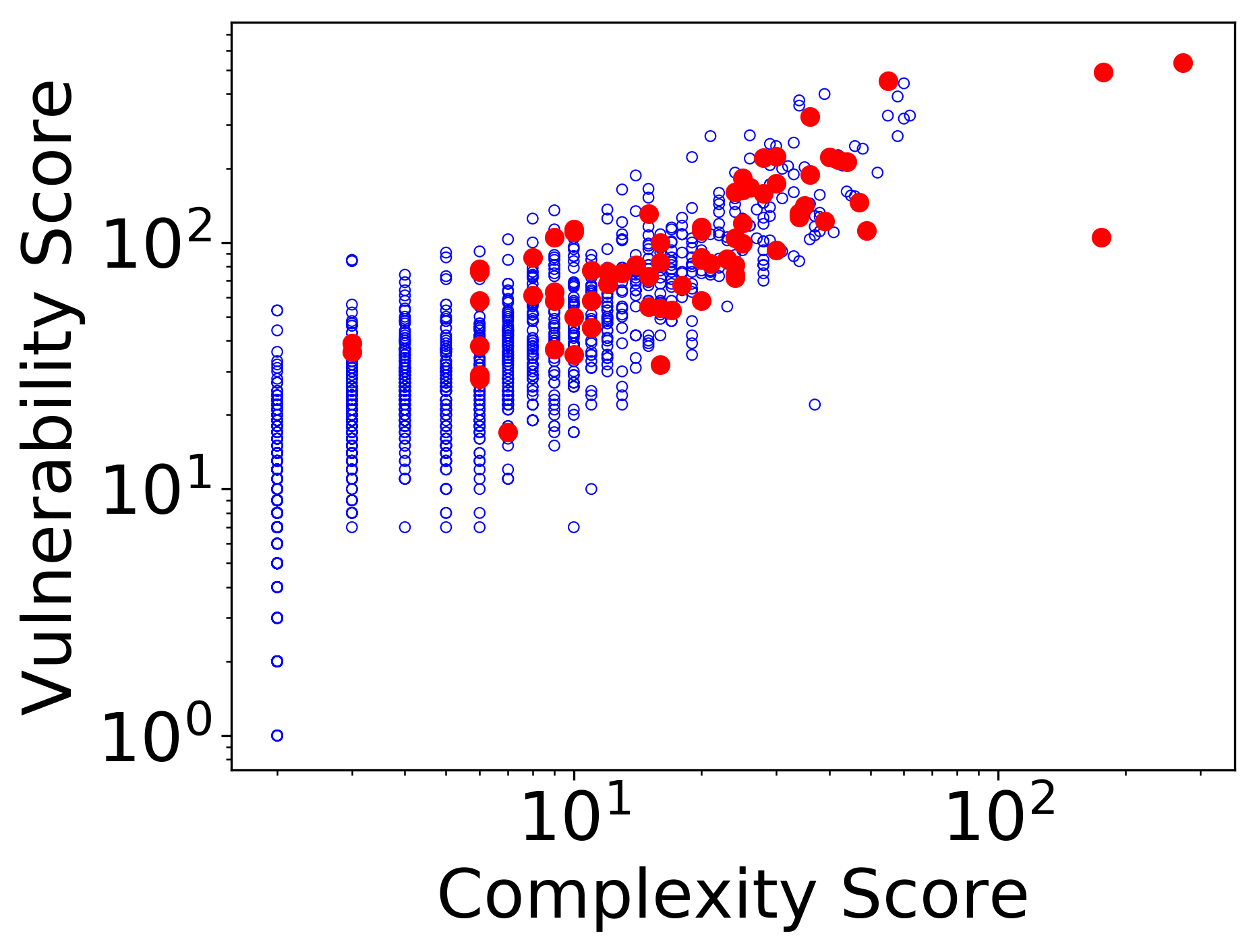}
\end{minipage}
\caption{\freetype}\label{fig:freetype}
\end{subfigure}
\begin{subfigure}[b]{0.19\textwidth}
\centering
\begin{minipage}{\linewidth}
\includegraphics[width=0.9\textwidth]{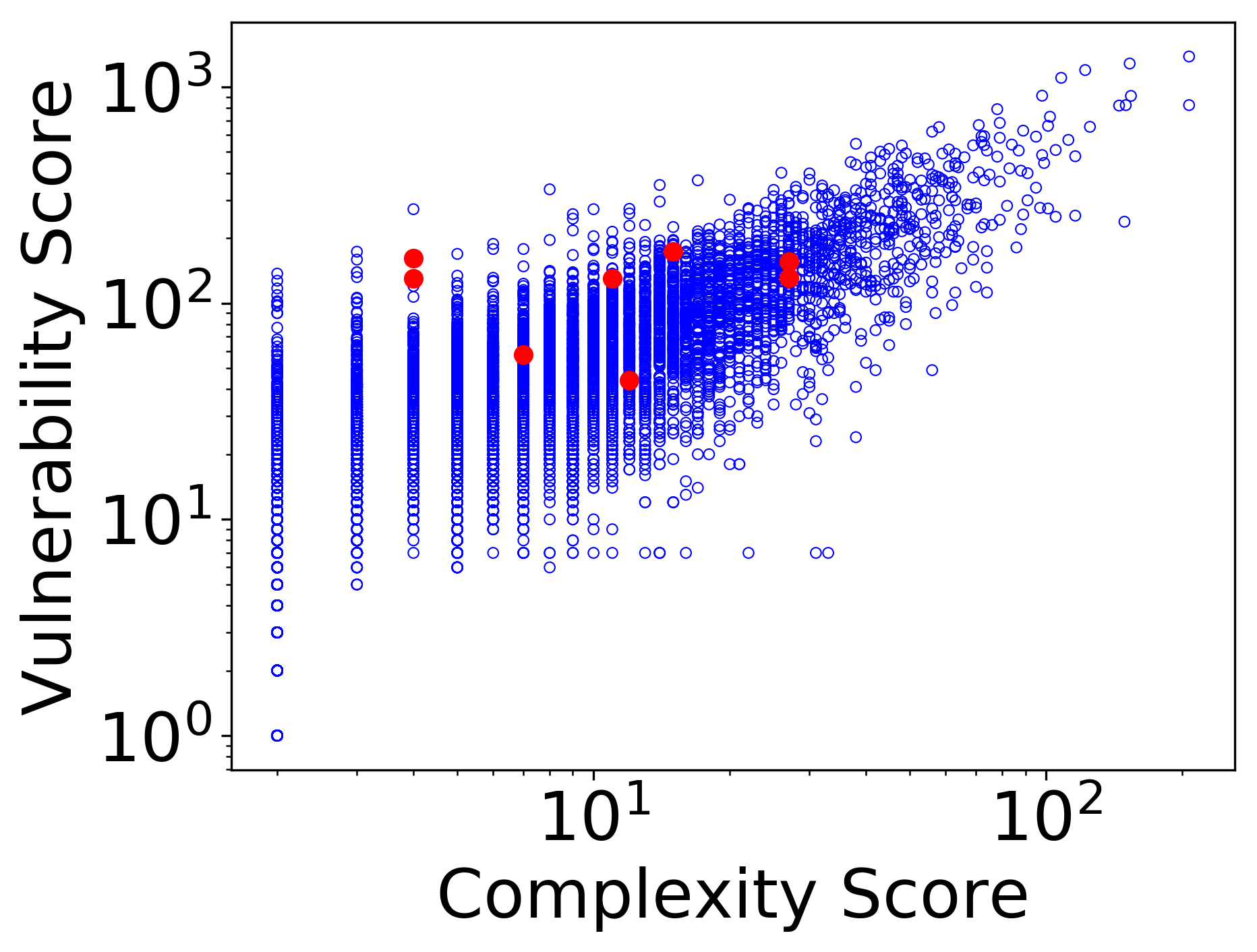}
\end{minipage}
\caption{\libav}\label{fig:libav}
\end{subfigure}
\begin{subfigure}[b]{0.19\textwidth}
\centering
\begin{minipage}{\linewidth}
\includegraphics[width=0.9\textwidth]{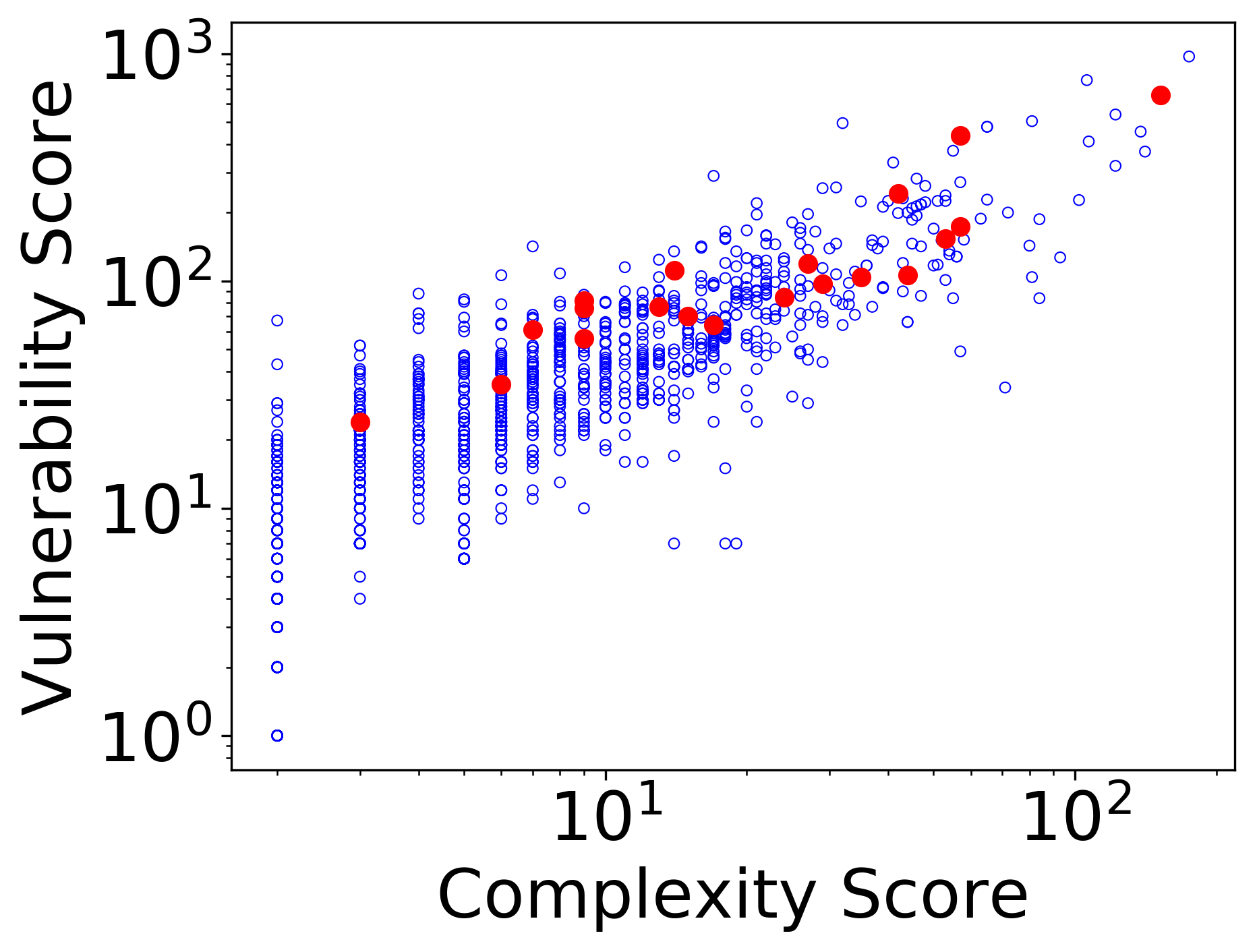}
\end{minipage}
\caption{\libtiff}\label{fig:libtiff}
\end{subfigure}

\begin{subfigure}[b]{0.19\textwidth}
\centering
\begin{minipage}{\linewidth}
\includegraphics[width=0.9\textwidth]{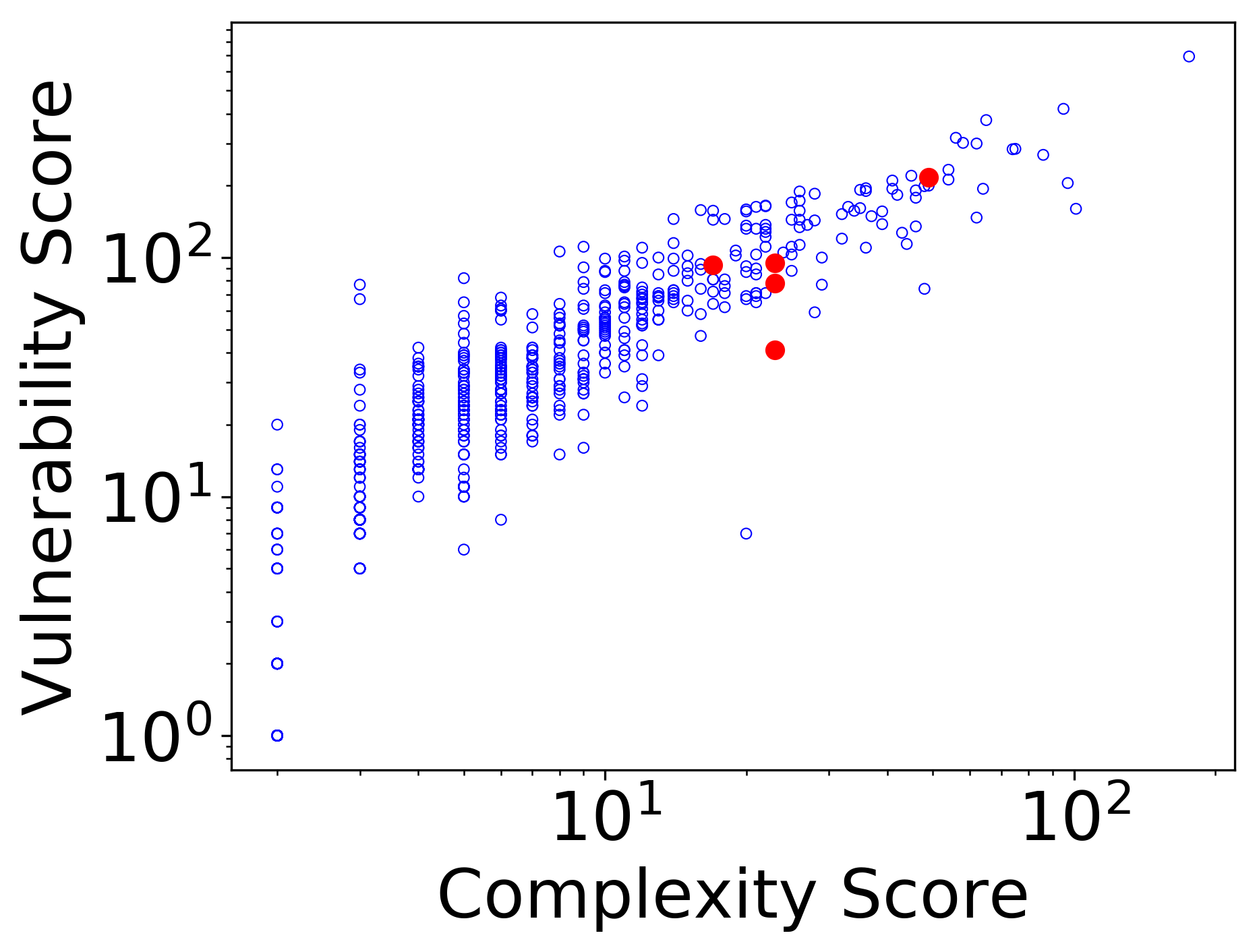}
\end{minipage}
\caption{\libxslt}\label{fig:libxslt}
\end{subfigure}
\begin{subfigure}[b]{0.19\textwidth}
\centering
\begin{minipage}{\linewidth}
\includegraphics[width=0.9\textwidth]{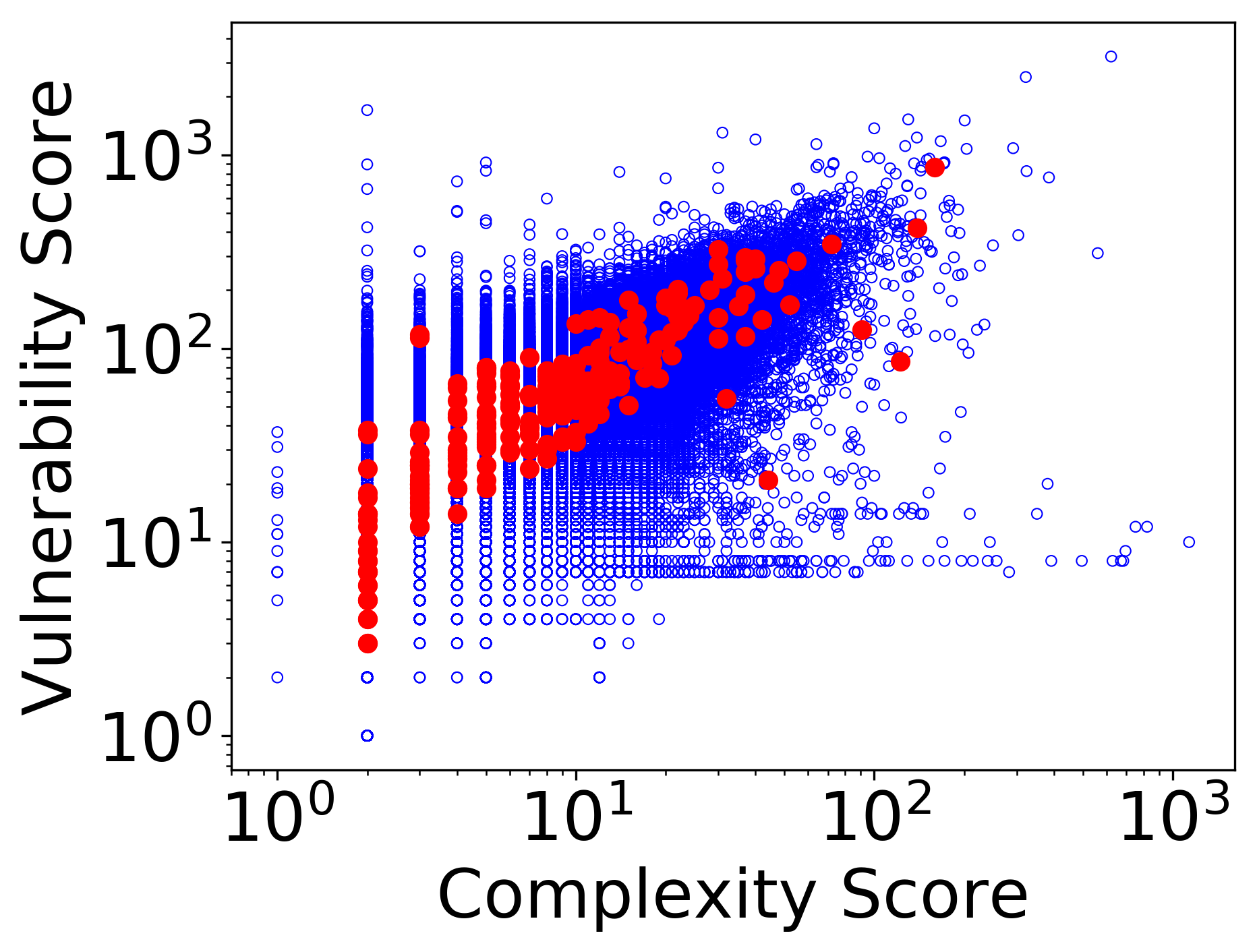}
\end{minipage}
\caption{\linux}\label{fig:linux}
\end{subfigure}
\begin{subfigure}[b]{0.19\textwidth}
\centering
\begin{minipage}{\linewidth}
\includegraphics[width=0.9\textwidth]{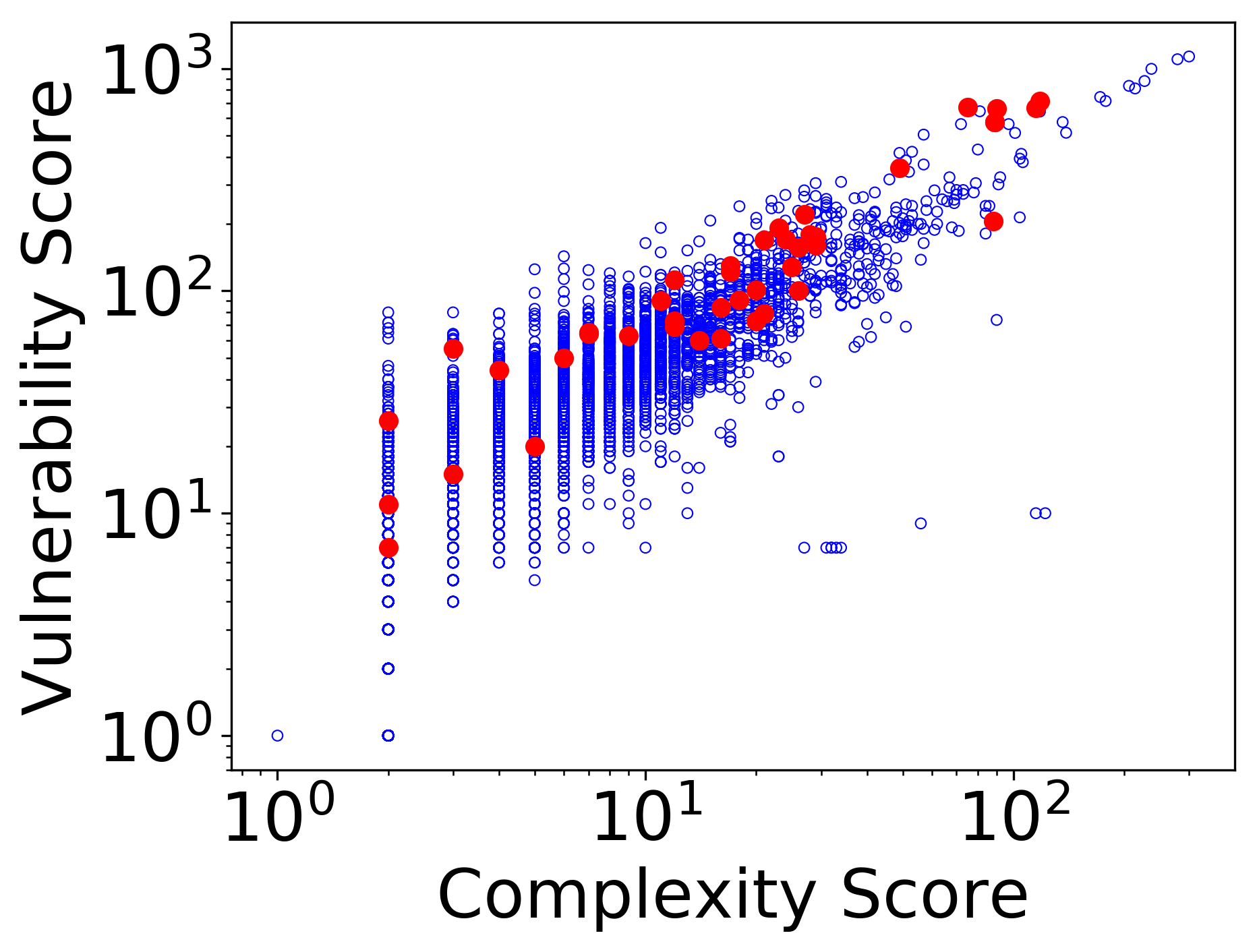}
\end{minipage}
\caption{\openssl}\label{fig:openssl}
\end{subfigure}
\begin{subfigure}[b]{0.19\textwidth}
	\centering
	\begin{minipage}{\linewidth}
		\includegraphics[width=0.9\textwidth]{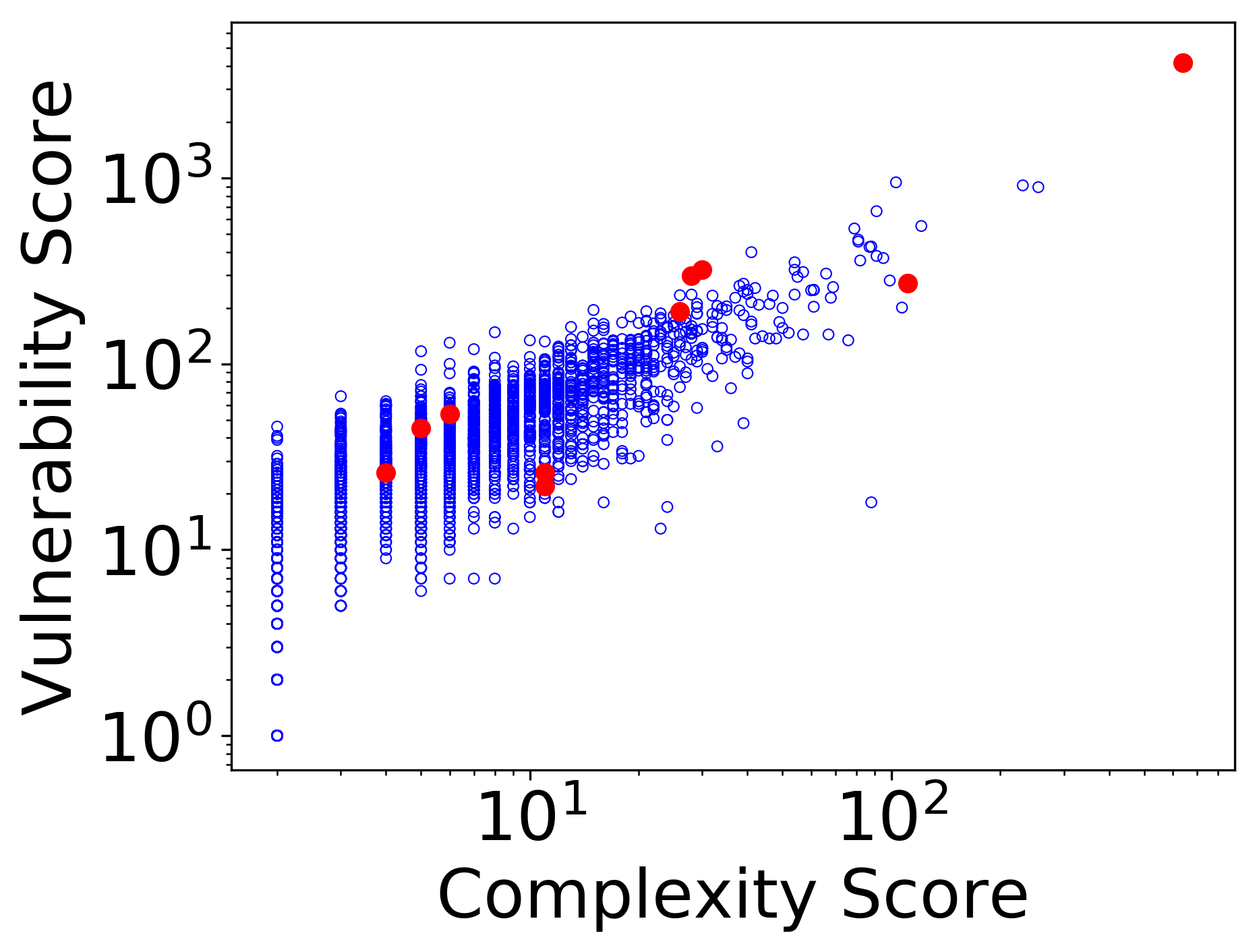}
	\end{minipage}
	\caption{\sqlite}\label{fig:sqlite}
\end{subfigure}
\begin{subfigure}[b]{0.19\textwidth}
	\centering
	\begin{minipage}{\linewidth}
		\includegraphics[width=0.9\textwidth]{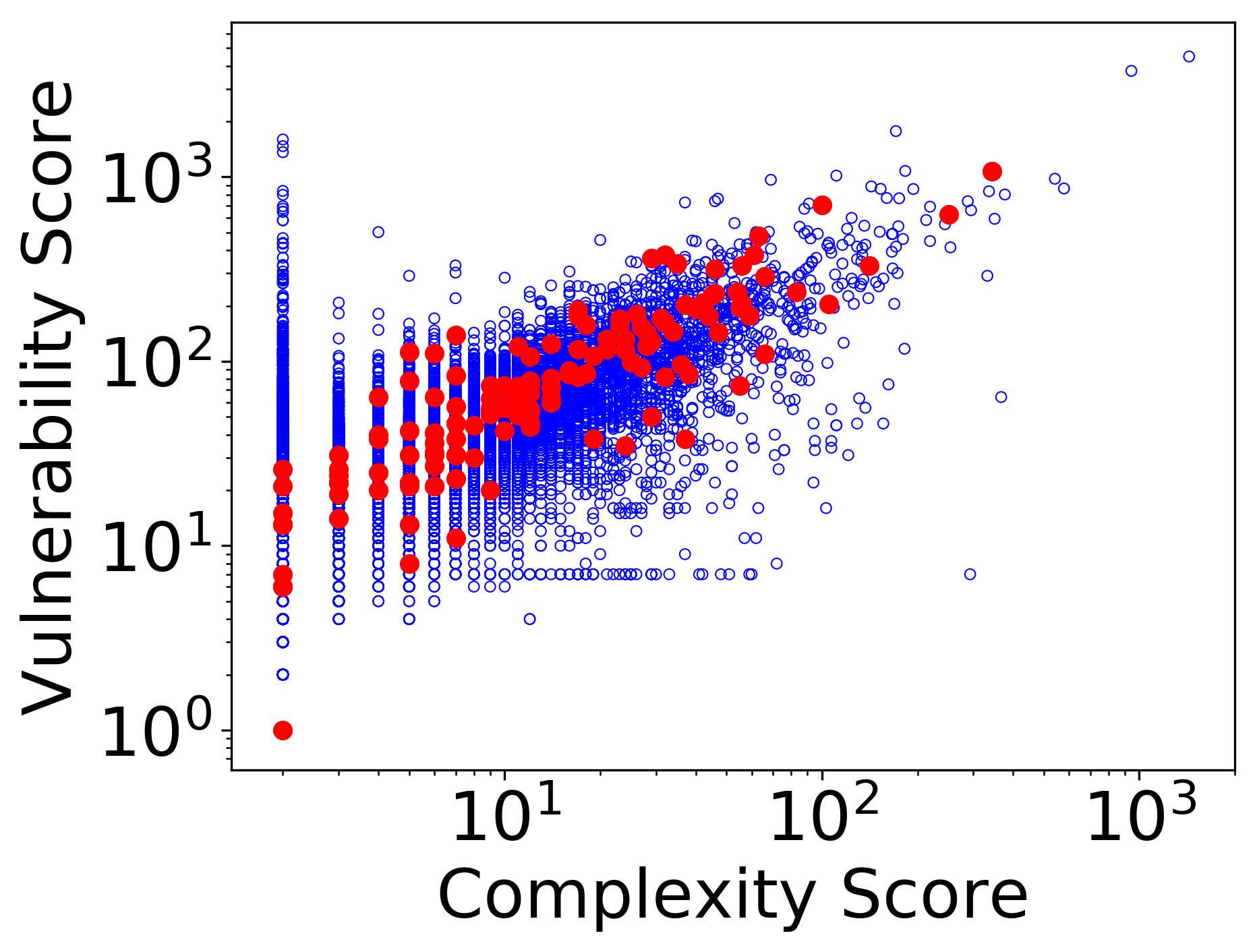}
	\end{minipage}
	\caption{\wireshark}\label{fig:wireshark}
\end{subfigure}
\vspace{-5pt}
\caption{Vulnerability Score vs. Complexity Score for Non-Vulnerable Functions (in Blue) and Vulnerable Functions (in Red)}\label{fig:rationality}
\end{figure*}

\noindent \textit{\textbf{Ground Truth.}} To obtain the ground truth for evaluating the effectiveness of \tool, we first manually identified the list of vulnerabilities that were disclosed before July 2018 in the \ning{11} projects from two vulnerability database websites: CVE Details \cite{cvedetails} and National Vulnerability Database \cite{NVD}, i.e., we collected all the vulnerabilities reported for the given version of the project from its release date to July 2018. \crc{CVEs in external libraries used in a project are not claimed to the project.}

The full~list~of~CVEs in most projects are recorded~by~the above two websites.
However, the patches of~the~CVEs~are~not well maintained and difficult to collect.
We obtained available patches of these CVEs in the 11 projects from an industrial collaborator, who offers vulnerability scanning services for C/C++ programs. Functions that are patched to fix the vulnerability are identified as vulnerable. The results are reported in the fourth and fifth columns of Table~\ref{table:project}.
As an example, we display the CVE list, available patches and corresponding patched functions of {\tt Libav}~at our website~\cite{Leopard}.

Some CVEs failed to be included in our research, as shown in the last column of Table \ref{table:project} because (i) there is no public detail about the fix that can directly identify the affected vulnerable functions as either the CVE affects some closed source projects or other reasons (e.g., CVE-2015-6607 and CVE-2015-5895 for {\tt \sqlite}); (ii) the fix does not involve direct code change on functions (e.g., CVE-2016-7958 for {\tt \wireshark} and CVE-2016-2183 for {\tt \openssl}).


\noindent\textit{\textbf{Research Questions.}} 
We designed the experiments to answer the following research questions:

\begin{itemize}[leftmargin=*]
\item {\bf Q1.} Is the binning step before the ranking step reasonable? (\S~\ref{sec:RatBR})
\item {\bf Q2.} Is our binning-and-ranking approach effective, and can it outperform baseline approaches, machine learning-based techniques and some off-the-shelf static scanners? (\S~\ref{sec:EffBR})
\item {\bf Q3.} What is the sensitivity of the metrics to the effectiveness of our framework? (\S~\ref{sec:sensM})
\item {\bf Q4.} What is the performance overhead (i.e., scalability) of our framework? (\S~\ref{sec:scale})
\item {\bf Q5.} What are the potential application scenarios of \tool? (\S~\ref{sec:evalapp})
\end{itemize}

\subsection{Rationality of Binning before Ranking (Q1)}\label{sec:RatBR}

To answer this question, we first computed the complexity score and vulnerability score, as in \S~\ref{sec:binning} and \S~\ref{sec:ranking}, for each function in all the projects (as shown in Table \ref{table:project}). Then we plotted in Fig. \ref{fig:rationality} the relationship between complexity score (i.e., $x$-axis) and vulnerability score (i.e., $y$-axis) using logarithmic scale, where vulnerable and non-vulnerable functions were respectively highlighted in red and blue. The result of {\tt BIND} is omitted for space limitations but is available on our website~\cite{Leopard}.

We can see from Fig. \ref{fig:rationality} that all projects share the similar patterns; vulnerable functions are scattered across non-vulnerable functions w.r.t. complexity score and vulnerability score; and there exists an approximately \crc{proportional relation} between complexity score and vulnerability score for vulnerable functions. Therefore, if we directly ranked the functions based on complexity metrics and/or vulnerability metrics, we would always favor those functions with high complexity score and high vulnerability score, and miss those with low-complexity but vulnerable (e.g., vulnerable functions located in the first 3 bins in Fig.~\ref{fig:binutils}, \ref{fig:linux} and \ref{fig:wireshark}).
Instead, by first binning the functions according to complexity score and then ranking the functions in each bin according to vulnerability score, our framework can effectively identify the potentially vulnerable functions at all levels of complexity (see details in \S~\ref{sec:EffBR}).
\crc{
For all 11 projects, the number of bins ranges from 56 to 206 with an average of 114.
Each bin has 301 functions on average, and 22\% of bins contain vulnerable functions.
Details of the function distribution among bins can be found at our website~\cite{Leopard}.
As can be seen from Fig.~\ref{fig:rationality}, bins with smaller complexity scores have more functions, and bins with larger complexity scores have more vulnerable functions.
Sparsity of bins with larger complexity scores benefits the selection of most vulnerable functions, while our ranking in bins with smaller complexity scores gives more chance to identify less complex but vulnerable functions.}
Moreover, Fig. \ref{fig:rationality} also visually indicates the severe imbalance between non-vulnerable and vulnerable functions (see the third and fourth columns of Table \ref{table:project}), which indicates traditional machine learning will over-fit and be less effective (more details will be discussed in \S~\ref{sec:EffBR}).

\begin{tcolorbox}[size=title, opacityfill=0.15]
Our binning-and-ranking approach is reasonable for predicting vulnerable functions at all levels of complexity.
\end{tcolorbox}

\begin{figure*}[!t]
\centering

\begin{subfigure}[b]{0.19\textwidth}
\centering
\begin{minipage}{\linewidth}
\includegraphics[width=0.9\textwidth]{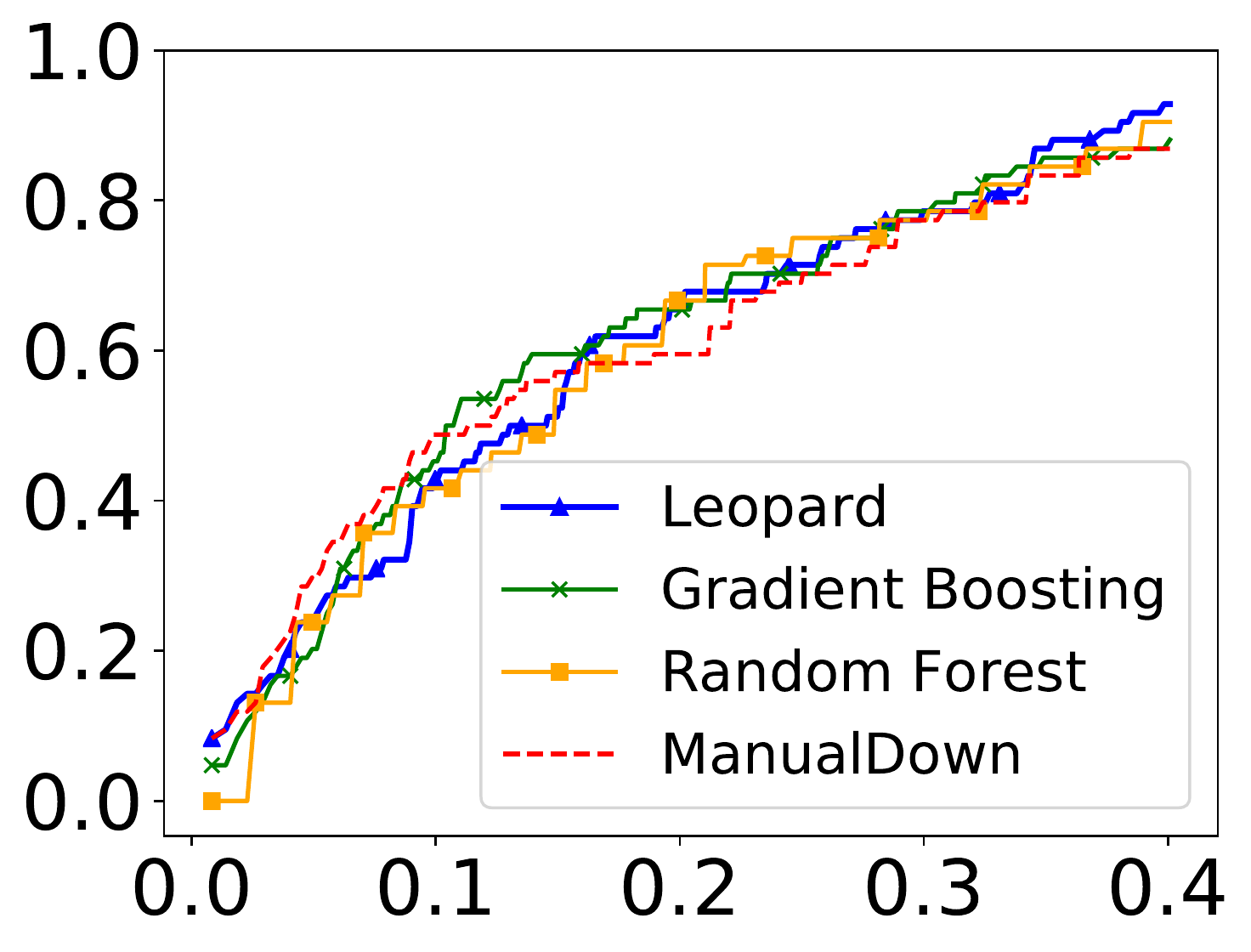}
\end{minipage}
\caption{\binutils}\label{fig:binutilsk}
\end{subfigure}
\begin{subfigure}[b]{0.19\textwidth}
\centering
\begin{minipage}{\linewidth}
\includegraphics[width=0.9\textwidth]{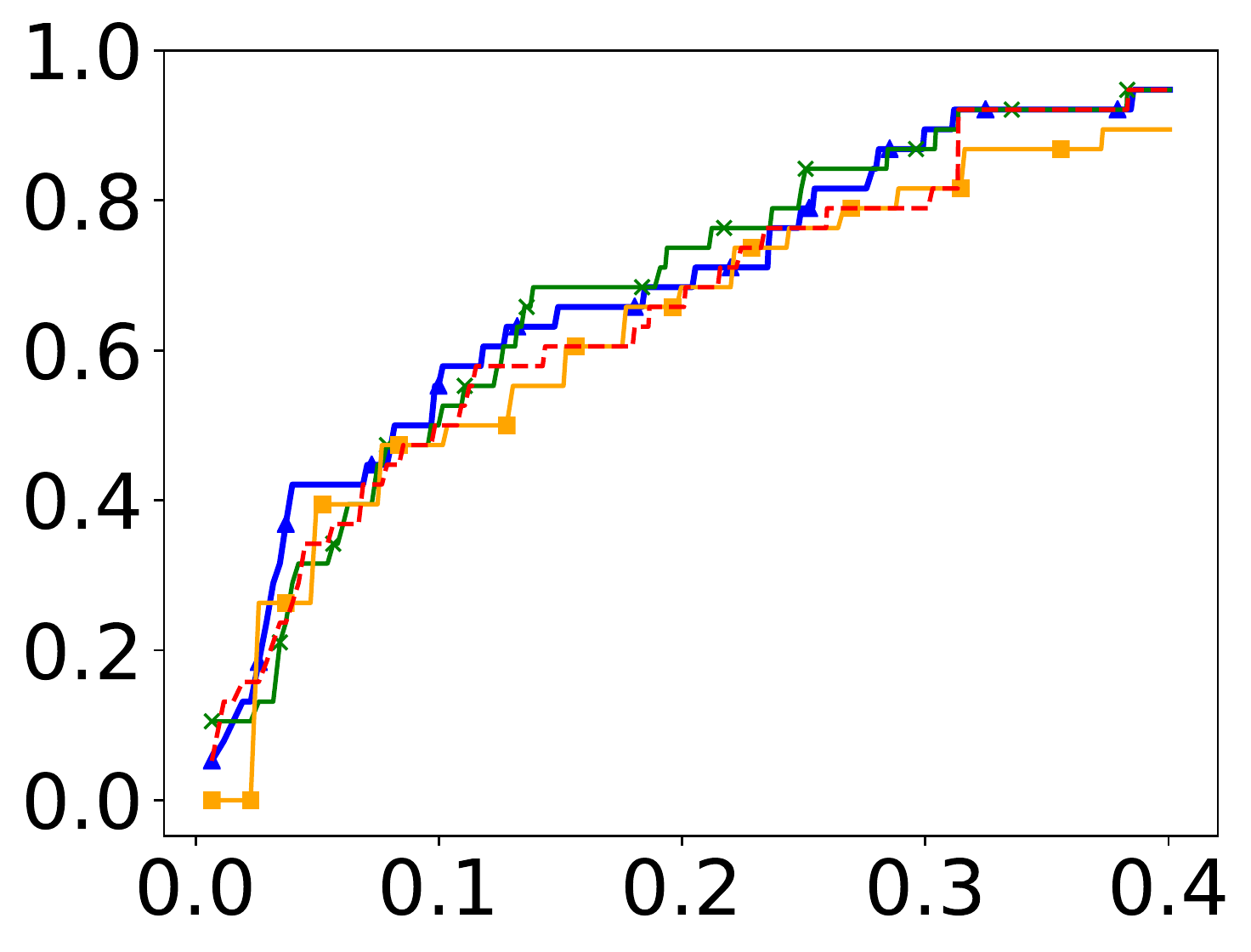}
\end{minipage}
\caption{\ffmpeg}\label{fig:QEMUk}
\end{subfigure}
\begin{subfigure}[b]{0.19\textwidth}
\centering
\begin{minipage}{\linewidth}
\includegraphics[width=0.9\textwidth]{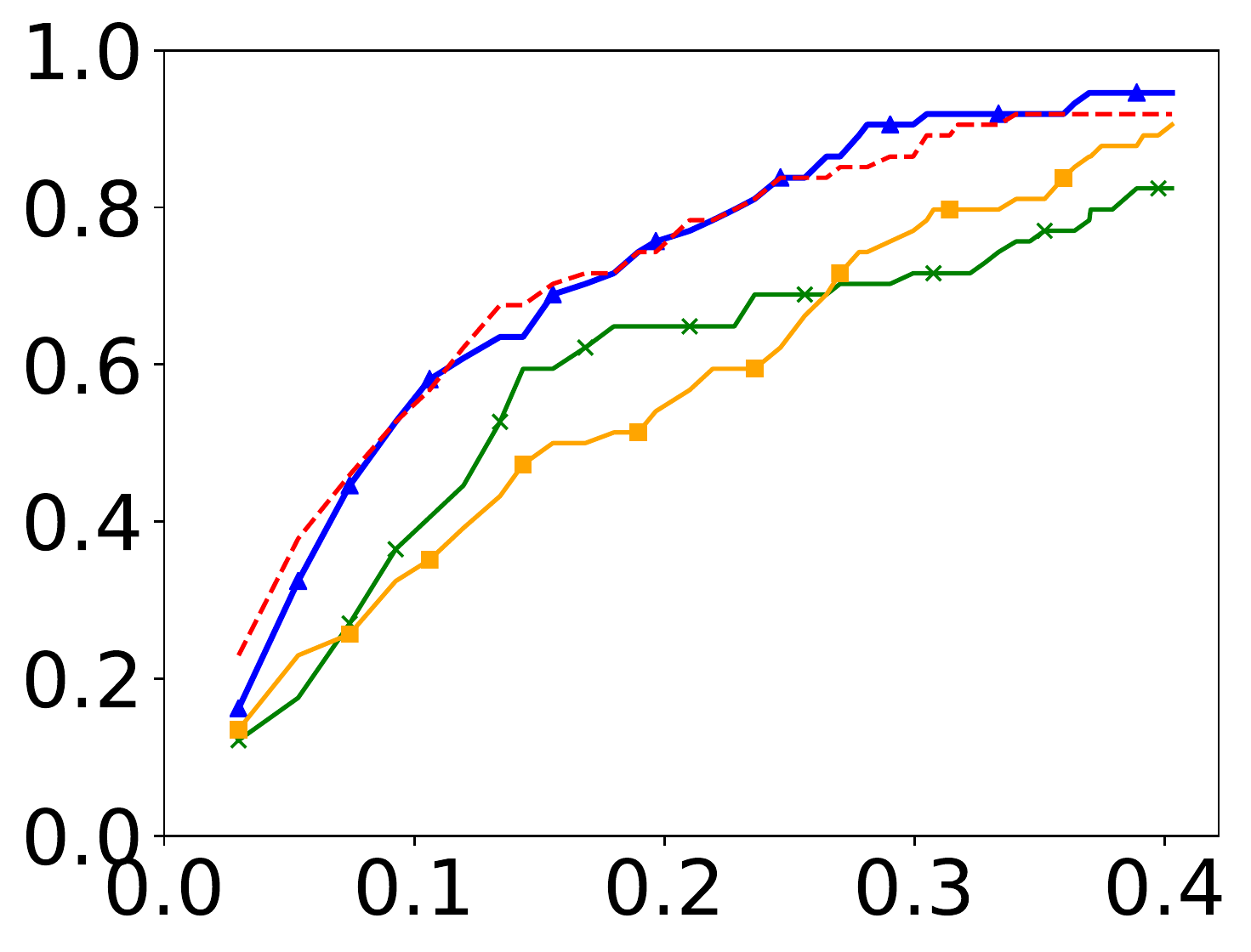}
\end{minipage}
\caption{\freetype}\label{fig:Asteriskk}
\end{subfigure}
\begin{subfigure}[b]{0.19\textwidth}
\centering
\begin{minipage}{\linewidth}
\includegraphics[width=0.9\textwidth]{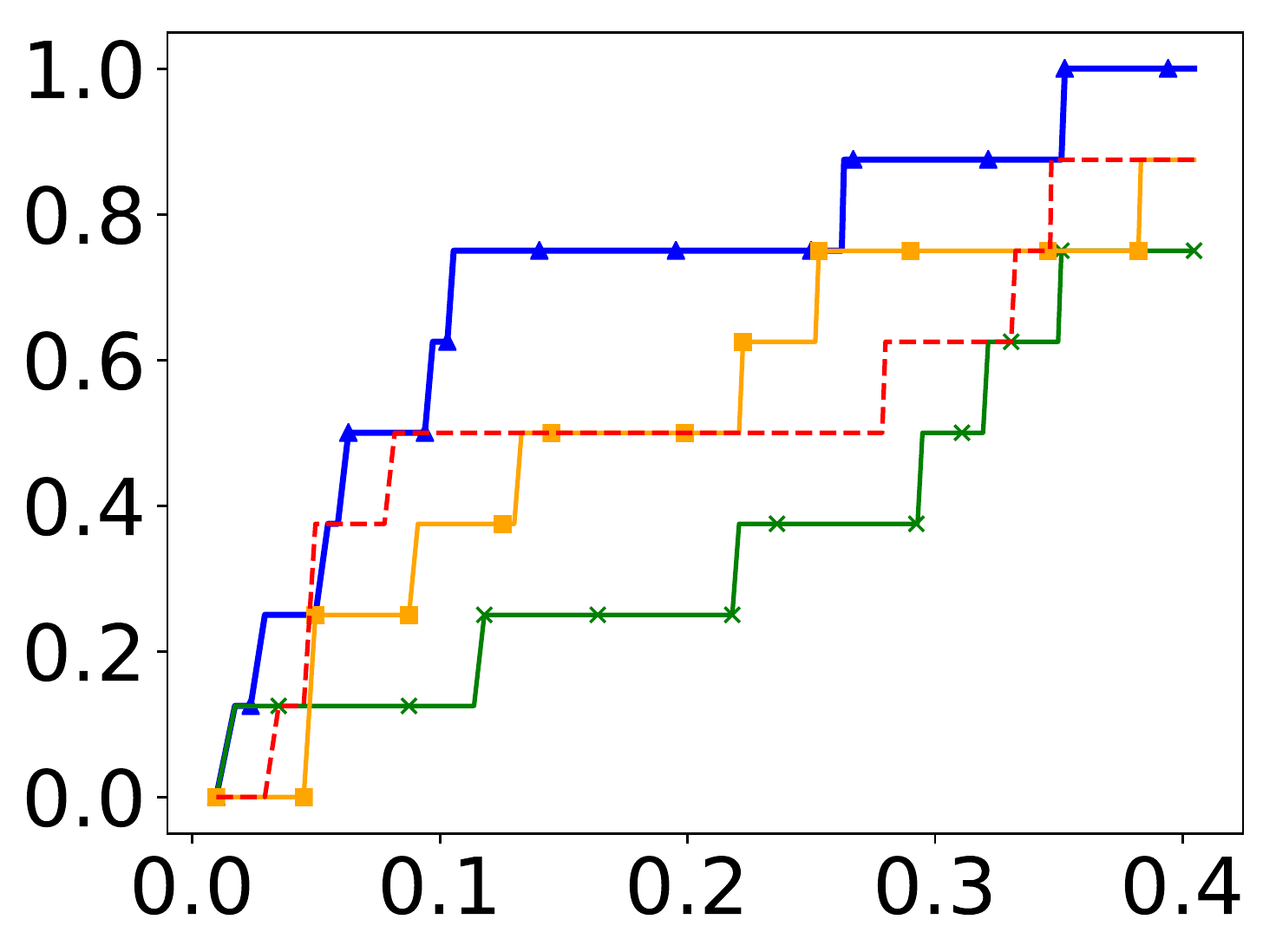}
\end{minipage}
\caption{\libav}\label{fig:libxml2k}
\end{subfigure}
\begin{subfigure}[b]{0.19\textwidth}
\centering
\begin{minipage}{\linewidth}
\includegraphics[width=0.9\textwidth]{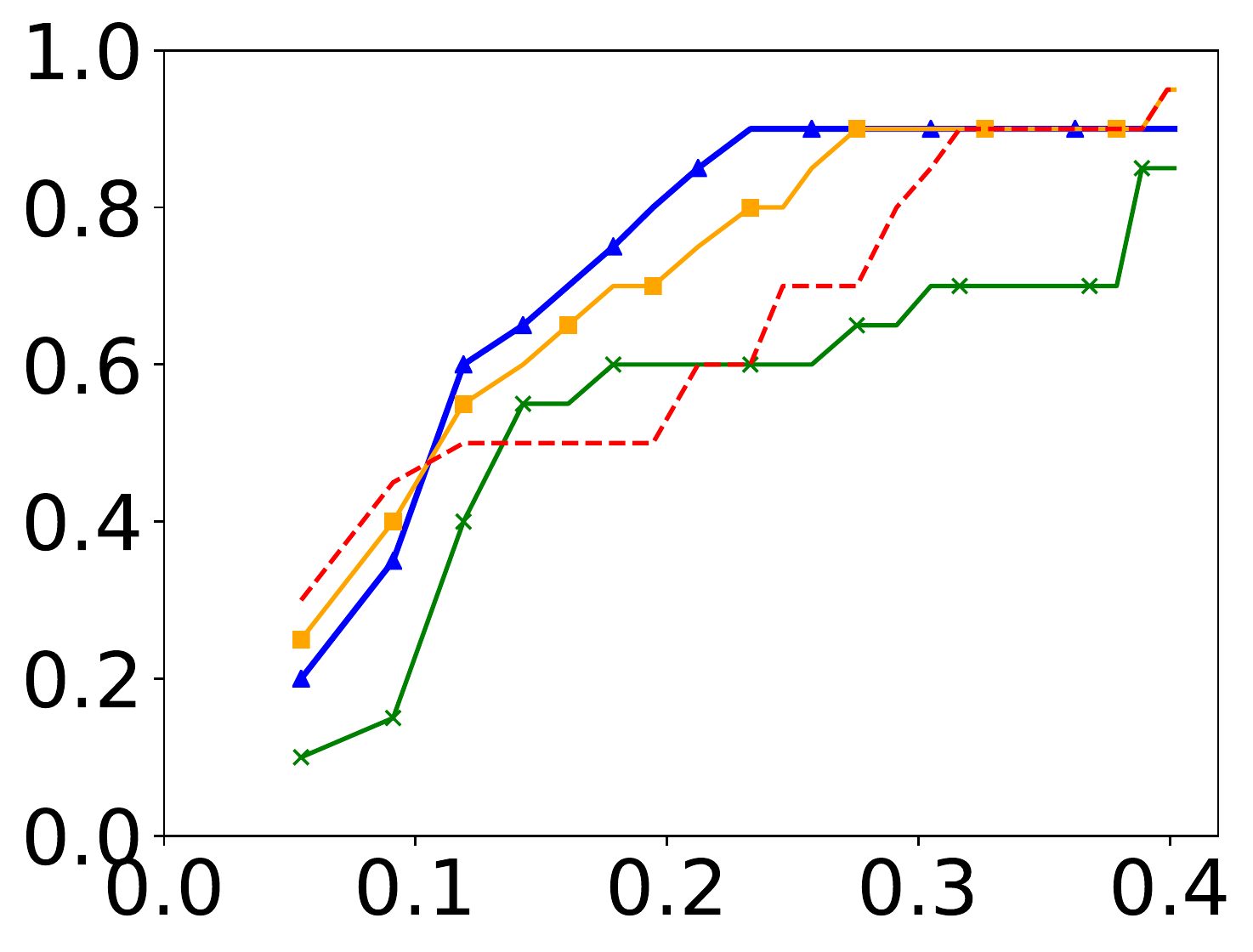}
\end{minipage}
\caption{\libtiff}\label{fig:libxsltk}
\end{subfigure}

\begin{subfigure}[b]{0.19\textwidth}
\centering
\begin{minipage}{\linewidth}
\includegraphics[width=0.9\textwidth]{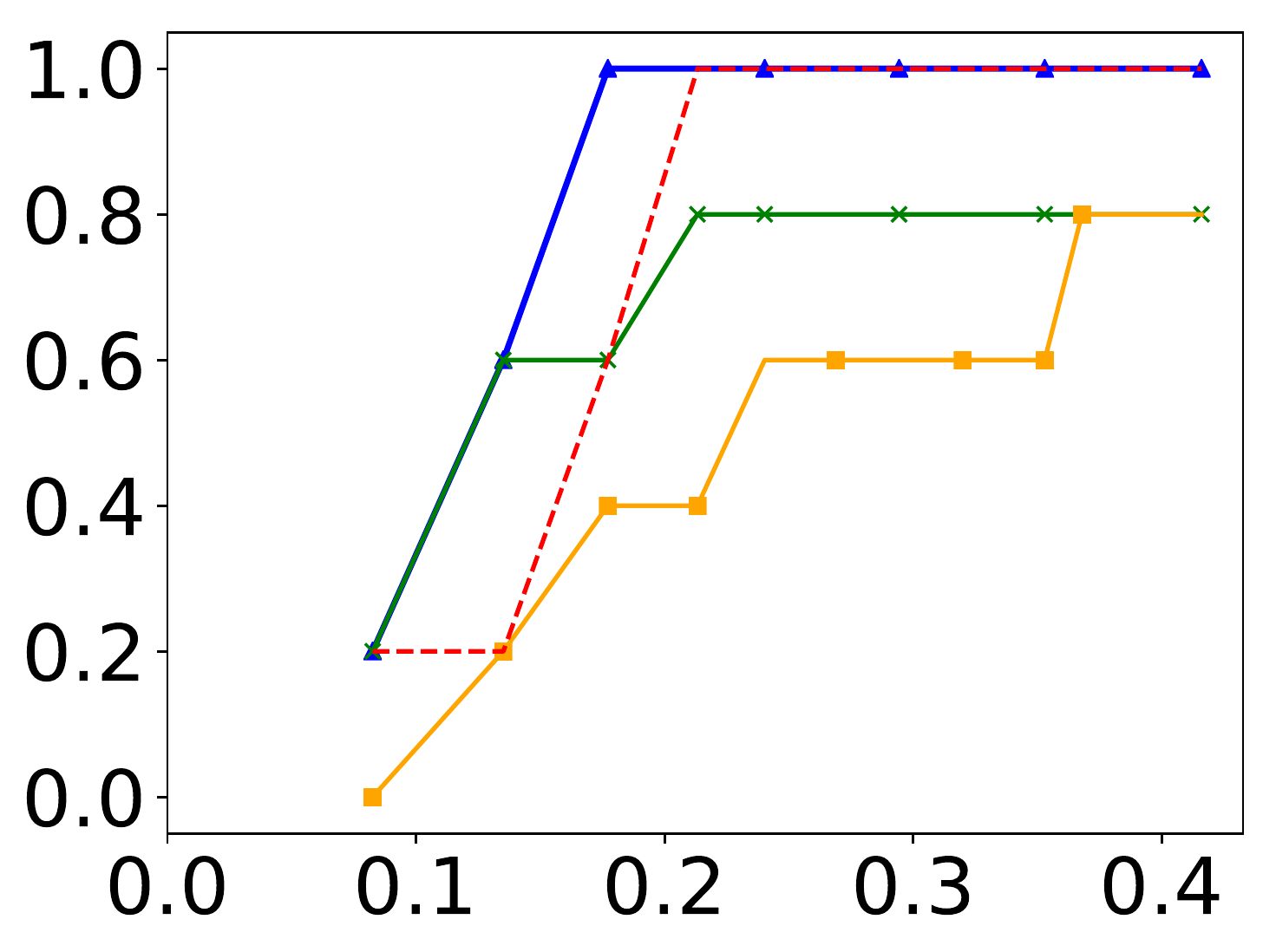}
\end{minipage}
\caption{\libxslt}\label{fig:OpenSSLk}
\end{subfigure}
\begin{subfigure}[b]{0.19\textwidth}
\centering
\begin{minipage}{\linewidth}
\includegraphics[width=0.9\textwidth]{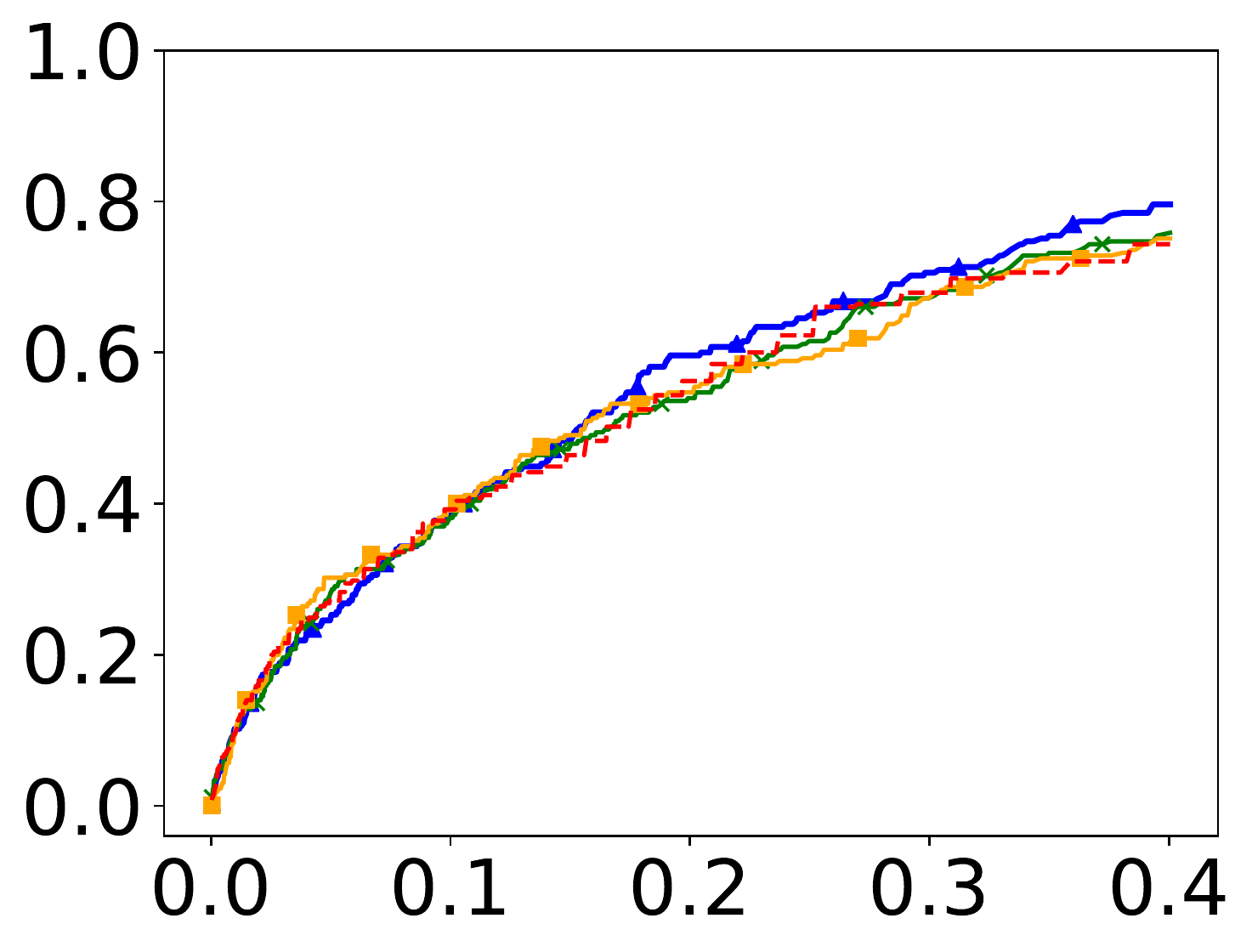}
\end{minipage}
\caption{\linux}\label{fig:tempk}
\end{subfigure}
\begin{subfigure}[b]{0.19\textwidth}
\centering
\begin{minipage}{\linewidth}
\includegraphics[width=0.9\textwidth]{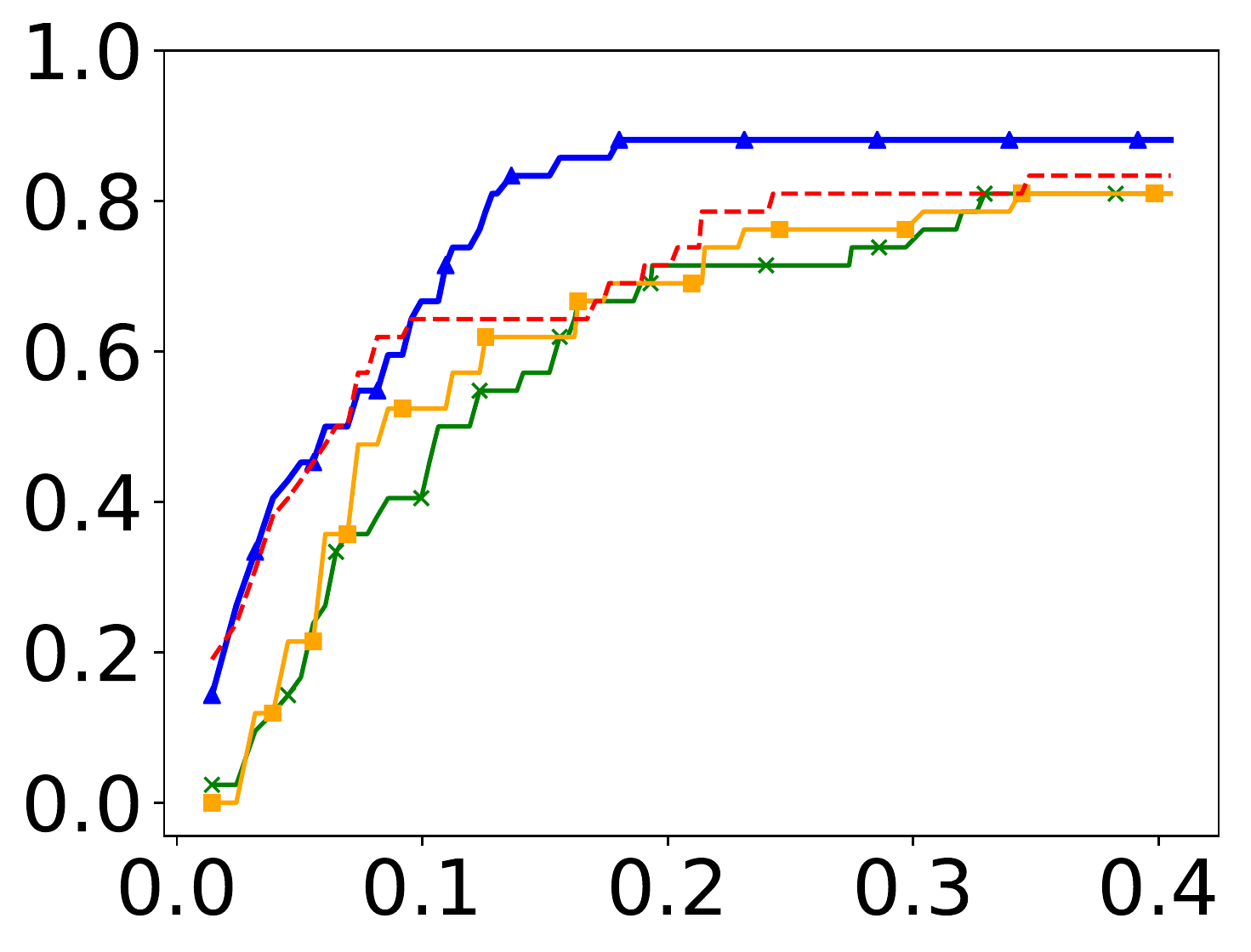}
\end{minipage}
\caption{\openssl}\label{fig:Linuxk}
\end{subfigure}
\begin{subfigure}[b]{0.19\textwidth}
	\centering
	\begin{minipage}{\linewidth}
		\includegraphics[width=0.9\textwidth]{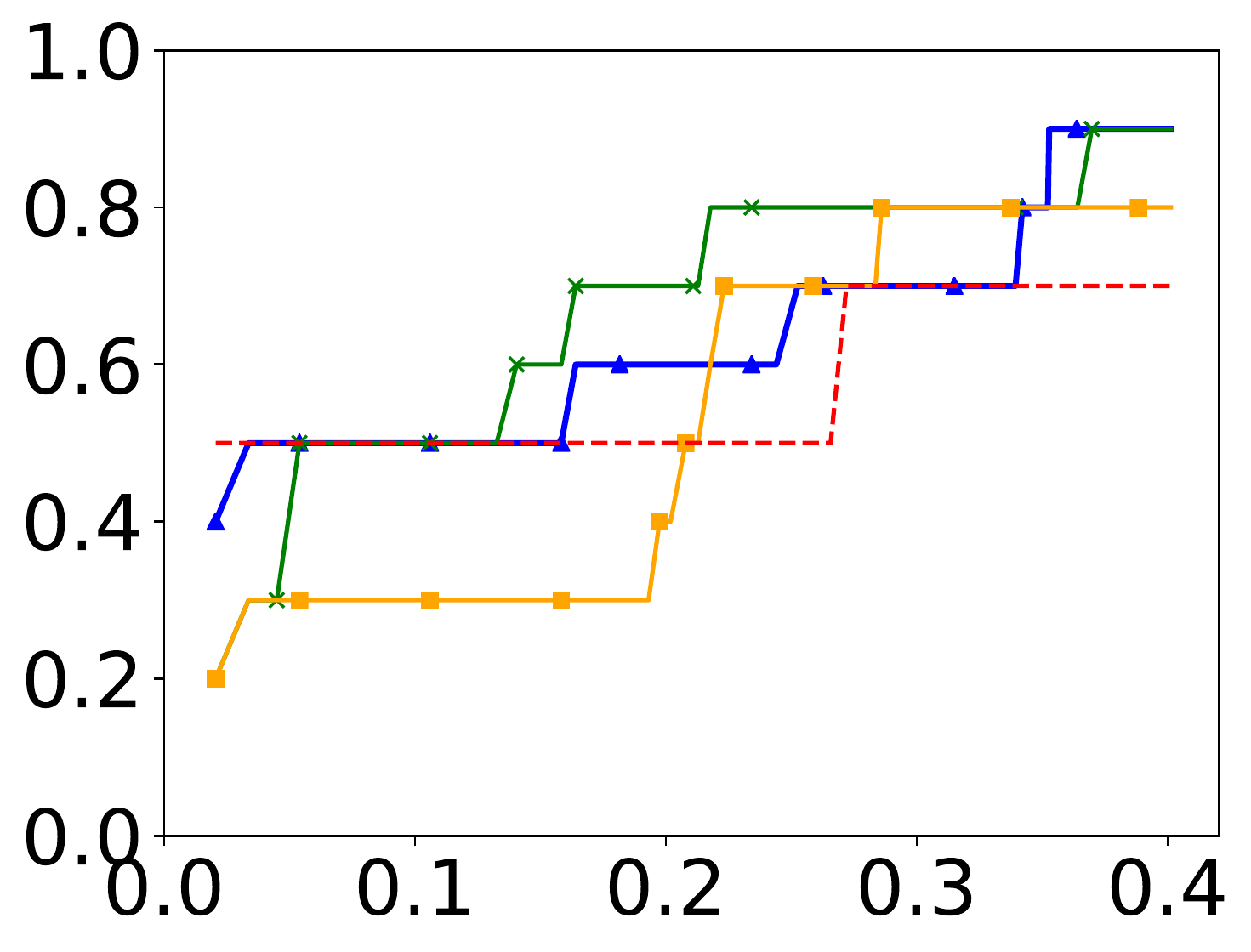}
	\end{minipage}
	\caption{\sqlite}\label{fig:sqlitek}
\end{subfigure}
\begin{subfigure}[b]{0.19\textwidth}
	\centering
	\begin{minipage}{\linewidth}
		\includegraphics[width=0.9\textwidth]{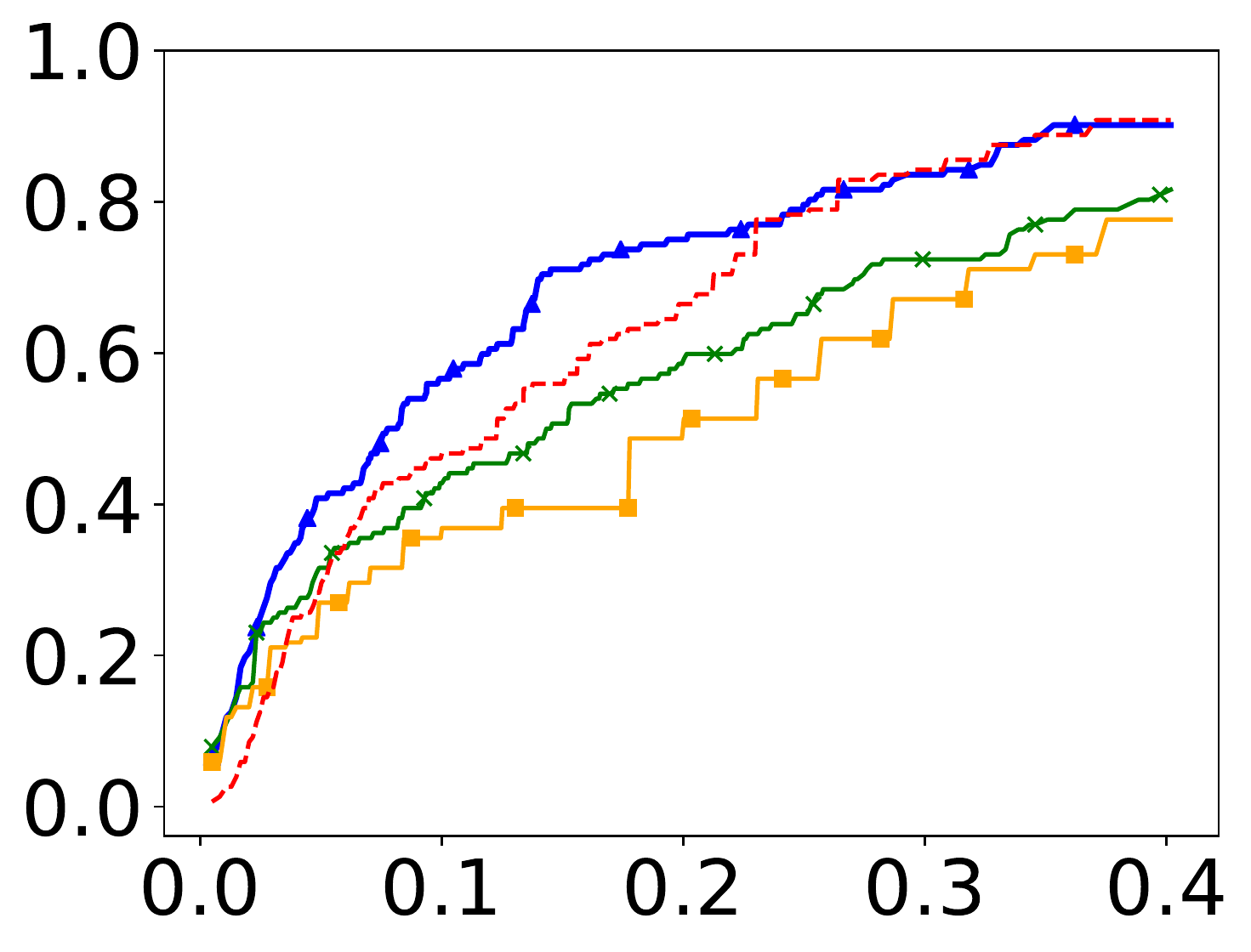}
	\end{minipage}
	\caption{\wireshark}\label{fig:Wiresharkk}
\end{subfigure}
\vspace{-5pt}
\caption{Percentage of Functions ({\it Iden. Func.}) that are Identified as Potentially Vulnerable, and Percentage of Vulnerable Functions ({\it Cov. Vul. Func.}) that are Covered by Those Identified Potentially Vulnerable Functions}\label{fig:effectiveBR}
\end{figure*}

\subsection{Effectiveness of Binning-and-Ranking (Q2)}\label{sec:EffBR}

We ran \tool on all the projects;
and analyzed its effectiveness when selecting different portion of functions, i.e., the parameter $p$ in the ranking step (see \S~\ref{sec:ranking}).
Here we used the percentage of functions (i.e., {\it Iden. Func.}) that are identified by \tool as potentially vulnerable, and the percentage of vulnerable functions (i.e., {\it Cov. Vul. Func.}) that are covered by those identified potentially vulnerable functions as the \ning{two} indicators of the effectiveness of our framework.
These \ning{two} indicators are used throughout the evaluation section.

The results are shown in Fig. \ref{fig:effectiveBR}, where the $x$-axis denotes {\it Iden. Func.} and the $y$-axis denotes {\it Cov. Vul. Func.}.
The legends are only shown in Fig. \ref{fig:binutilsk} and omitted in others for clarity; and the result of {\tt BIND} is omitted but available on the website~\cite{Leopard}. In general, as {\it Iden. Func.} increases, the indicator {\it Cov. Vul. Func.} also increases. For a small value (e.g., 20\%) of {\it Iden. Func.}, our binning-and-ranking approach can achieve a high value for {\it Cov. Vul. Func.} (e.g., 74\%).
Furthermore, we also report
how many \ning{vulnerable functions} are covered when we identify certain percentage of functions as vulnerable in Table \ref{table:vfcve}.
When identifying 5 \%, 10\%, 15\%, 20\%, 25\% and 30\% of functions as vulnerable, we can cover \ning{29\%, 49\%, 64\% 74\%, 78\% and 85\%} of vulnerable functions. 
This means by identifying a small part of functions as vulnerable, we cover a large portion of vulnerable functions,
 which can narrow down the assessment space for security experts.

\begin{table}[t]
	\scriptsize
	\centering
	\caption{Identified and Covered Vulnerable Functions}\label{table:vfcve}
	\vspace{-5pt}
	\begin{tabular}{|c||c|c|c|>{\bfseries}c|c|c|}
		\hline
		\multirow{2}{*}{Project} & \multicolumn{6}{c|}{Iden. Func. (\%)} \\\cline{2-7}
		{} & 5\% & 10\% & 15\% & 20\% & 25\% & 30\% \\\hline
		\bind & 55.6 & 55.6 & 66.7 & 66.7 & 66.7 & 88.9 \\\hline
		\binutils & 23.8 & 42.9 & 51.2 & 65.5 & 71.4 & 78.6 \\\hline
		\ffmpeg & 42.1 & 55.3 & 65.8 & 68.4 & 78.9 & 89.5 \\\hline
		\freetype & 16.2 & 52.7 & 63.5 & 75.7 & 83.8 & 90.5 \\\hline
		\libav & 25.0 & 62.5 & 75.0 & 75.0 & 75.0 & 87.5 \\\hline
		\libtiff & 0.0 & 35.0 & 65.0 & 80.0 & 90.0 & 90.0 \\\hline
		\libxslt & 0.0 & 20.0 & 60.0 & 100.0 & 100.0 & 100.0 \\\hline
		\linux & 24.9 & 38.9 & 48.3 & 59.6 & 64.9 & 70.6 \\\hline
		\openssl & 42.9 & 66.7 & 83.3 & 88.1 & 88.1 & 88.1 \\\hline
		\sqlite & 50.0 & 50.0 & 50.0 & 60.0 & 60.0 & 70.0 \\\hline
		\wireshark & 40.8 & 56.6 & 71.1 & 75.0 & 79.6 & 83.6 \\\hline
		Average & 29.2 & 48.7 & 63.6 & 74.0 & 78.0 & 85.2 \\\hline

	\end{tabular}
\end{table}

\noindent\textit{\textbf{Comparison to Baseline Approaches.}} A recent study~\cite{zhou2018far} on 42 existing cross-project defect prediction models and two state-of-the-art unsupervised defect prediction models~\cite{nam2015clami,zhang2016cross} has indicated that, simply ranking functions based on source lines of code (SLOC) in an increasing (i.e., ManualUp) or decreasing (i.e., ManualDown) order can achieve comparable or even superior prediction performance compared to most defect prediction models. We put the results of ManualUp (which is much worse than \tool) at our website~\cite{Leopard}, and only show results of ManualDown in this section.

\begin{table}[t]
	\scriptsize
	\centering
	\caption{Comparison of \tool to Existing Approaches}
	\label{table:avrrecall}
	\vspace{-5pt}
	\begin{tabular}{|c|c|c|c|c|c|c|}
		\hline
		\multirow{2}{*}{Approach} & \multicolumn{6}{c|}{Iden. Func. (\%)}   \\ \cline{2-7}
		& 5\%  & 10\% & 15\% & 20\% & 25\% & 30\% \\ \hline\hline
		\tool & 29.2 & \textbf{48.7} & \textbf{63.6} & \textbf{74.0} & \textbf{78.0} & \textbf{85.2} \\ \hline
		ManualDown                & \textbf{34.3} & 47.9 & 54.4 & 63.7 & 70.6 & 78.2 \\ \hline
		Random Forest             & 25.8 & 37.7 & 48.8 & 58.8 & 68.7 & 75.6  \\ \hline
		Gradient Boosting         & 22.1 & 39.3 & 54.4 & 60.9 & 67.8 & 73.0 \\ \hline
		
	\end{tabular}
\end{table}

In Fig.~\ref{fig:effectiveBR}, the comparison of {\it Cov. Vul. Func.} between \tool and ManualDown is shown for each project.
\tool shows better results for all projects except for {\tt Binutils} and {\tt FreeType}, where both approaches have similar performance.
On average, compared to ManualDown, 9.2\%, 10.3\% and 7.4\% improvement are achieved when identifying 15\%, 20\% and 25\% of functions as vulnerable, as shown in Table~\ref{table:avrrecall}; and we identify 15.6\%, 13.8\% and 11.8\% less codes (measured in SLOC) than ManualDown.
\crc{On average, 96.8\% of ManualDown's true positives are covered by \tool}. 
This demonstrates the effectiveness of \tool.

\noindent\textit{\textbf{Comparison to Machine Learning-Based Techniques.}} We also conducted experiments to compare our framework with four machine learning-based techniques, namely random forest (RF), gradient boosting (GB), naive Bayes (NB) and support vector classification (SVC). The four techniques used all \todo{4} complexity metrics and \todo{11} vulnerability metrics as the features, and conducted a cross-project prediction by first training a model with the data from ten of the 11 projects and then using the model to predict the probability of being vulnerable for the functions in the other one project. By rotating the project to predict, we obtained the prediction results for all 11 projects. A larger predicted probability indicates that a function is more likely vulnerable. We rank the functions according to the probabilities, and identify a list of high-probability functions as vulnerable. A fair comparison to \tool can be drawn when the same number of functions is identified. The results are shown in Fig.~\ref{fig:effectiveBR} and Table~\ref{table:avrrecall}.

As shown in Fig.~\ref{fig:effectiveBR}, an obvious shortcoming of RF and GB is the unstable performance among different projects.
It indicates that machine learning approaches highly depend on the large knowledge base of various vulnerable functions, which are however hard to obtain.
Specifically, RF only shows similar or slightly better performance than \tool in Fig.~\ref{fig:binutils} and~\ref{fig:ffmpeg}, while GB only shows similar performance in Fig. \ref{fig:binutils}, \ref{fig:ffmpeg} and \ref{fig:sqlite}.
\tool outperforms RF and GB in Fig. \ref{fig:freetype}, \ref{fig:libav}, \ref{fig:libtiff} \ref{fig:libxslt}, \ref{fig:linux}, \ref{fig:openssl} and \ref{fig:wireshark}.
Both RF and GB performs even worse than the ManualDown baseline in Fig. \ref{fig:freetype}, \ref{fig:openssl} and \ref{fig:wireshark}.
As numerically shown in Table~\ref{table:avrrecall}, when identifying 20\% of functions, RF and GB separately cover 15.2\% and 13.1\% less of ground truth than \tool. Again, \tool does not rely on any prior knowledge about a large set of vulnerabilities but machine learning-based techniques do. NB and SVC presented extremely lower recalls among the four typical machine learning algorithms.
Hence, we omitted the results and put them at our website~\cite{Leopard}.
\crc{Note that 11 projects~may not be an adequate dataset for training and testing, especially given the severe imbalance between vulnerable and non-vulnerable functions, the validity of conclusions drawn can be threatened. However, such a prerequisite for prior knowledge of vulnerable functions motivate our design of \tool.}

\noindent\textit{\textbf{Comparison to Static Scanners.}}
We also applied two popular static software scanner tools to investigate their vulnerability prediction capability on our dataset, including an open source tool, Cppcheck~\cite{Cppcheck}, and a commercial tool.
To avoid legal disputes, we hide the name of the commercial one and refer it as S***.
Cppcheck and S*** are among the most popular static code analysis tools used to detect bugs and vulnerabilities in software.
Both tools report the suspicious vulnerable statements. Whenever an alarm locates within the vulnerable functions in our ground truth, we claim a true positive for that tool.
The number of total alarms reported by these two tools and the recall can be found in Table~\ref{table:scanner}.
Cppcheck was able to analyze all 11 projects and identified a few vulnerable functions in {\tt Binutils}, {\tt FreeType} and {\tt Wireshark}.
S*** failed to analyze {\tt Linux}; and for the other 10 projects only a few vulnerable functions are detected in {\tt LibTIFF}.
\crc{Static scanners often rely on very concrete vulnerability patterns. Subtle pattern mismatch will cause false positives and negatives. Thus. their recalls are nearly 0, which indicate that they are not promising for general vulnerability identification.}

\begin{table}[t]
    \scriptsize
	\centering
	\caption{Number of Alarms and Recall of Static Scanners}
	\begin{tabular}{|c||c|c|c|c|}
		\hline
		\multirow{2}{*}{Project}  & \multicolumn{2}{c|}{S***} & \multicolumn{2}{c|}{Cppcheck} \\ \cline{2-5}
		& \#Alarm        & Recall       & \#Alarm        & Recall       \\ \hline \hline
		\bind      & 250            & 0.0          & 45             & 0.0          \\ \hline
		\binutils  & 106            & 0.0          & 261            & 0.012        \\ \hline
		\ffmpeg    & 42             & 0.0          & 306            & 0.0          \\ \hline
		\freetype  & 0              & 0.0          & 82             & 0.054        \\ \hline
		\libav     & 19             & 0.0          & 138            & 0.0          \\ \hline
		\libtiff   & 76             & 0.1          & 10             & 0.0          \\ \hline
		\libxslt   & 20             & 0.0          & 6              & 0.0          \\ \hline
		\linux     & -              & -            & 3864           & 0.0          \\ \hline
		\openssl   & 76             & 0.0          & 33             & 0.0          \\ \hline
		\sqlite    & 20             & 0.0          & 37             & 0.0          \\ \hline
		\wireshark & 0              & 0.0          & 115            & 0.007        \\ \hline
	\end{tabular}
	\label{table:scanner}
\end{table}

\noindent \textit{\textbf{False Negative Analysis.}} By examining the vulnerable functions that \tool fails to cover when 40\% functions are identified, we summarize three main reasons for false negatives: 1) they are involved in some logical vulnerabilities which are hard to be revealed by function metrics; 2) they are implicated via some fixes indirectly related to the CVE, e.g., if a fix changes the function signature, callers of this function should not be counted as vulnerable; or 3) security critical information is in their surrounding context and unseen from the function itself, e.g., calculation of complicated pointer offsets sometime is done via a separate function, where no pointer metrics can be inferred, thus resulting in a lower vulnerability score.
For the first case, such vulnerabilities are generally hard to identify via static analysis, and should not be a concern of our approach.
Case two is also irrelevant to the validity of our approach.
A mitigation for the third case is to include taint information to our vulnerability metrics, as will be discussed in \S\ref{sec:discuss}.


\noindent \textit{\textbf{False Positive Analysis.}} 
Balancing the generality, accuracy and scalability is always a very challenging task for static analysis.
Since \tool is designed to reveal general vulnerabilities, it is impossible to avoid false positives.
However, \tool aims to assist vulnerability assessment rather than a stand alone static analysis tool. False analysis is therefore not a critical criteria for evaluating its capability.
Furthermore, some vulnerabilities are previously patched in history, secretly patched
~\cite{Xu2017} or currently unexposed, 
and it is impossible to confirm whether they are indeed false positives.
This is also reflected in the experiments in \S~\ref{sec:evalapp}, where new vulnerabilities have been found in the reported potential vulnerable functions.

\begin{tcolorbox}[size=title, opacityfill=0.15]
Our binning-and-ranking approach is effective, i.e., identifying \ning{20\%} of functions as vulnerable to cover \ning{74.0\%} of vulnerable functions on average. Such a small portion of functions can be very useful for security experts, as will be shown in our application of \tool in \S~\ref{sec:evalapp}. Besides, \tool outperforms machine learning-based techniques and static analysis-based approaches.
\end{tcolorbox}

%% file: sec0401-evaluation.tex
\subsection{Sensitivity of the Metrics (Q3)}\label{sec:sensM}

\begin{figure}[!t]
	\centering
	\includegraphics[width=0.5\textwidth]{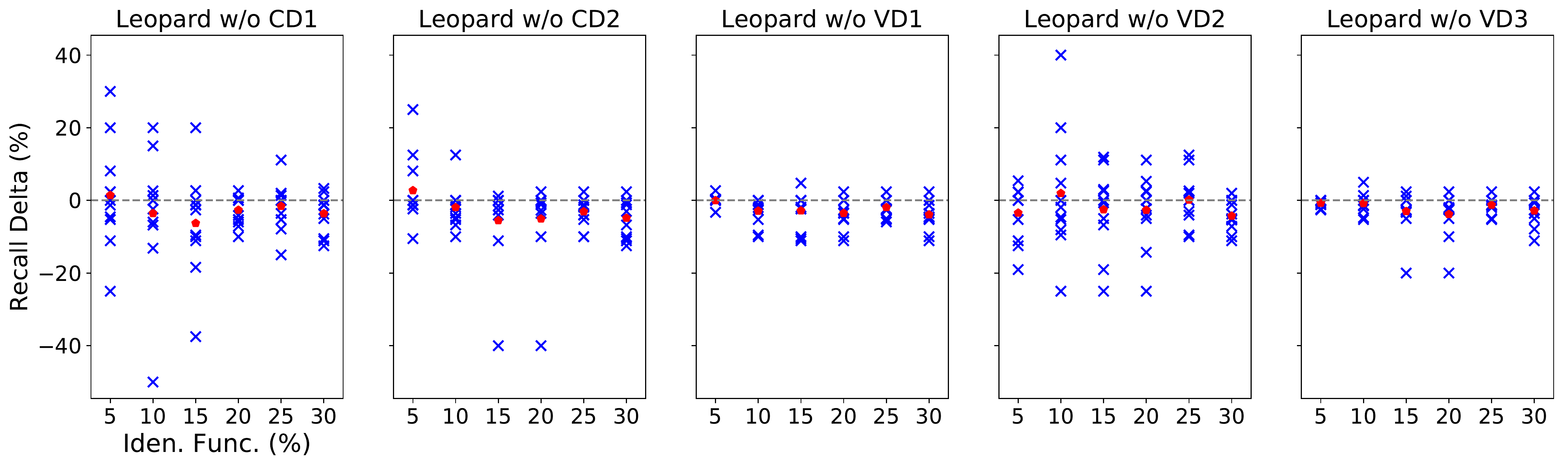}
	\vspace{-10pt}
	\caption{Sensitivity Analysis Results of Metrics}
	\label{fig:sensitivity1}
\end{figure}

To evaluate the sensitivity of the complexity and vulnerability~metrics to our framework, we removed one of the dimensions of the complexity and vulnerability metrics from \tool, and~then~ran \tool on all the projects. 
We show the sensitivity results of complexity metrics and vulnerability metrics in Fig.~\ref{fig:sensitivity1}.
The $x$-axis and $y$-axis represent {\it Iden. Func.} and \ning{the {\it delta} of recall (i.e., {\it Cov. Vul. Func.}) compared to \tool with all metrics}.
After removing one dimension of metrics, the recall delta of each project when identifying certain percentage of functions are labeled by blue cross marks, where positive delta means improvement in performance, and negative ones means degradation. The red dots are average recall delta among all 11 projects.

We can see from Fig.~\ref{fig:sensitivity1} that, basically, there are much more degradation than improvement when removing any dimension of metrics.
Moreover, the average recall deltas across projects are negative for {\it Iden. Func.} at 15\%, 20\%, 25\%, and 30\% in all five experiments, i.e., less vulnerable functions are covered when the same percentage of functions is identified as vulnerable.
Some improvement of average recall delta at 5\% and 10\% actually results from some relatively large improvements of only a few projects.
Specifically, most significant degradation occurs when the cyclomatic complexity metrics (i.e., CD1) is removed, and most significant average degradation occurs when the loop structure metrics are removed, which indicates they make substantial contribution to our framework.
It also proves the necessity of our binning strategy.
With the above observation, we can conclude that all dimensions of our complexity and vulnerability metrics contribute to the effectiveness of \tool, but complexity metrics contribute the most;
and  it is difficult or even impossible to derive an optimal model for the metric combination that can work well for all ranges of {\it Iden. Func.} for all projects.
Hence, we design a generic but not optimal model that treats each metric equally.

%
\begin{tcolorbox}[size=title, opacityfill=0.15]
Complexity metrics~significantly contribute to \tool; and it is difficult to derive~an~optimal metric model that works for all projects, which motivates our generic model without sacrificing much effectiveness.
\end{tcolorbox}

%
%
%

\subsection{Scalability of Our Framework (Q4)}\label{sec:scale}

To evaluate the scalability of our framework, we collected~the~time of extracting complexity and vulnerability metrics 
and the time of identifying potentially vulnerable functions by \tool. The detailed results~are reported at our website~\cite{Leopard}.
The time used to build the code property graph and query the graph to obtain metric values depends on the number of functions in each project. For small-scale projects, it respectively takes 2 and 45 minutes to build and query the graph; and it takes hours for large-scale projects (i.e., {\tt Wireshark} and {\tt Linux}).
It takes less than 50 seconds to identify 100\% functions even for {\tt Linux}. These results demonstrate that our framework scales well for large-size projects like {\tt Linux}.
For machine learning-based techniques, GB on average takes 9 minutes to train the model and make the prediction for each project, and RF takes 5 minutes.
Considering they also depend on the metrics calculation, \tool is more efficient.
S*** basically takes several minutes to finish the static analysis but requires the project to be well compiled and built, and fail to handle {\tt Linux}.
The lightweight static scanner Cppcheck shows comparable performance as \tool.

\begin{tcolorbox}[size=title, opacityfill=0.15]
Our framework scales well and can be applied to large-scale applications like {\tt Linux}.
\end{tcolorbox}

%% file: sec0403-evaluation-manual-auditing.tex
\subsection{Application of \tool (Q5)}\label{sec:evalapp}

\noindent \textit{\textbf{Manual Auditing.}}
Code review is a popular approach for vulnerability hunting. 
In this section, we demonstrate the role that \tool plays in helping security experts to hunt vulnerabilities with a case study of \ffmpeg.
In order not to overwhelm the security expert, we showed the top 1\% candidates with \tool, which is a list of 128 functions with detailed complexity and vulnerability metric scores, as well as the specific variables involved in the metrics, e.g., the variables involved in control predicates.
The security expert is experienced with code review and is familiar with the basic implementation and code structures of {\it FFmpeg}.
He firstly grouped the functions into different modules and chose {\it libavformat} as the target, which is the module responsible for the streaming protocols and conversion,
and has been prone to vulnerabilities in history.
Among all 128 functions, 13 of them are in {\it libavformat}.
He spent one day to find a {\it divide-by-zero} bug in one of the functions, with CVE-2018-14394 assigned.
Intuitively, he thinks the maximum of data-dependent control structures metrics (with the variables involved) more interesting, as he can be guided to trace backward and/or forward the data flow of these sensitive variables.
Detailed discussion about the aforementioned case can be found at our website~\cite{Leopard}.

%% file: sec0404-evaluation-fuzzing.tex

\noindent \textit{\textbf{Directed Fuzzing.}}
As discussed in \S~\ref{sec:application}, \tool can supply targets for directed fuzzing.
Experimentally, we ran \tool on {\tt PHP} {\tt 5.6.30} (a popular general-purpose scripting language that is especially suited to web development) and identified around 500 functions as potentially vulnerable.
Notice that {\tt PHP} is used by more than 80\% of all the websites, and {\tt 5.6.30} is the~current stable version.
Thus {\tt PHP} is well-tested~by its users, developers, and security researchers, and it is difficult to find vulnerabilities.
We selected top 500 functions reported by \tool as the target sites for Hawkeye for bug hunting.
We divided {\tt PHP} into several modules based on its architecture and focused on the functions in the modules (e.g., mbstring and Zend) that are related to file system and network data as they are often reachable through entry points.
We excluded the functions in those well-fuzzed modules (e.g., SQLite, phar and gd). This manual filtering process is different from manual auditing as the security expert does not pinpoint the vulnerability directly.
After 6-hour fuzzing, we discovered~six vulnerabilities in {\tt PHP} {\tt 5.6.30} with details shown in Table~\ref{table:cves}.

\begin{table}[t]
	\scriptsize
	\centering
	\caption{Zero-Day Vulnerabilities in PHP}\label{table:cves}
	\vspace{-5pt}
	\begin{tabular}{|c|c|c|cc|}
		\hline
		\multirow{2}{*}{Module} & \multirow{2}{*}{CVE-ID} & \multirow{2}{*}{Type}   & \multicolumn{2}{c|}{Reproducible?}  \\
		\cline{4-5}
		&          &               & 32-bit & 64-bit                    \\
		\hline\hline
		php::mbstring    & CVE-2017-9224                            & stack out-of-bound read                   &     \cmark   &    \cmark            \\
		php::mbstring    & CVE-2017-9225                            & heap out-of-bound write                   &       \cmark &         \xmark                  \\
		php::mbstring    & CVE-2017-9226                           & heap out-of-bound write                   &       \cmark &          \cmark                 \\
		php::Zend        & CVE-2017-9227                          & stack out-of-bound read                     &       \cmark &          \cmark                 \\
		php::mbstring    & CVE-2017-9228                           & heap out-of-bound write                   &     \cmark   &          \cmark                 \\
		php::mbstring    & CVE-2017-9229                           & invalid dereference DoS                   &       \cmark &           \cmark                \\
		\hline
	\end{tabular}
\end{table}

\noindent \textit{\textbf{Seed Prioritization.}}
In \S~\ref{sec:application}, we also discussed the application of applying the results of \tool to the seed evaluation process during fuzzing.
We used \tool to generate function level scores for three real-world open-source projects and utilized the scores to provide guidance to FOT~\cite{Chen:2018:FVC:3236024.3264593}.
The three projects are \emph{mjs}~\cite{mjs} (a Javascript engine for embedded systems), \emph{xed}~\cite{pin} (the disassembler used in \emph{Intel-Pin}) and \emph{radare2}~\cite{radare2} (a popular open source reverse engineering framework).
For the experiment purpose, we ran FOT with and without the guidance from \tool for 24 hours and collected the detected crashes.

Table~\ref{table:crashes} shows the detailed performance differences of FOT with and without \tool.
From the results, \tool can help FOT to detect 127\% more crashes in 24 hours on average.
\crc{Finally, seven new bugs are found in \emph{mjs}, seven new bugs are found in \emph{xed}, and a new vulnerability (CVE-2018-14017) is exposed in \emph{radare2}.}

\begin{tcolorbox}[size=title, opacityfill=0.15]
These results showed that \tool can substantially enhance the vulnerability finding for a limited time budget, which is the original purpose of designing \tool.
\end{tcolorbox}

\begin{table}[t]
	\scriptsize
	\centering
	\caption{Crashes Detected in 24 Hours by FOT with and without the Results from \tool}\label{table:crashes}
	\vspace{-5pt}
\begin{tabular}{|l|c|c|c|c|}
	\hline
	Project    & mjs & xed & radare2 & \textit{Average} \\ \hline\hline
	w/o \tool & 181 & 720 & 7 & 303 \\ \hline
	with \tool  & \textbf{251} & \textbf{1800} & \textbf{9} & \textbf{687} \\ \hline
\end{tabular}

\end{table}

%% file: sec05-discuss.tex

\section{Metrics Extension}\label{sec:discuss}

The set of complexity and vulnerability metrics can~be~refined and extended, to highlight interesting functions via~capturing different perspectives. To this end, we have identified~the~following information to be vital to further improve our findings.

\noindent \textit{{\textbf{Taint Information.}}} Leveraging taint information will help an analyst to identify the functions that process the external (i.e., taint) input. In general, functions that process or propagate the taint~information can be considered quite interesting for further assessment. Hence, incorporating the taint information into vulnerability metrics will further enhance the \tool's ranking step by assigning more weight (or importance) to the functions that process or propagate the taint information.


\noindent \textit{{\textbf{Vulnerability History.}}} In general, when a vulnerability~is~reported, the functions related to the vulnerability will go through an intensive security assessment during the patching process. Hence, such information can be used to refine the ranking by either: (1) giving more importance to recently patched~functions
due to the verified reachability, with considerable risks of incomplete patch or introducing new issues, or (2) giving~low~priority~to~the~functions that are patched long before the release of the current version, assuming that the functions have gone through a thorough security assessment and it is not worth the effort to re-assess~it. 

\textit{{\textbf{Domain Knowledge.}}} Domain knowledge can play a vital~role in prioritizing the interesting functions for further assessment. Information such as the modules that are currently fuzzed~by others or the knowledge about the modules that are shared by two or more projects can be used to refine \tool's ranking. 

%% file: sec06-related-work.tex
\section{Related Work}\label{sec:related}

Here we discuss the most closely related work that 
aim at assisting security experts during vulnerability assessment.

\noindent\textit{\textbf{Pattern-Based Approaches.}} Pattern-based approaches use patterns of known~vulnerabilities~to identify potentially vulnerable code. 
Initially, code scanners~(e.g., Flawfinder~\cite{Flawfinder}, PScan~\cite{PScan}, RATS~\cite{RATS} and~ITS4 \cite{Viega2000}) were proposed~to match vulnerability patterns. 
These scanners are efficient and practical, but fail to identify complex vulnerabilities as the patterns are often coarse-grained and straightforward. Differently, our approach does not require any patterns or prior knowledge of vulnerabilities.

Since then, security researchers have started to leverage more advanced static analysis techniques for pattern-based vulnerability identification (e.g., \cite{Larus2004,Son2011,Yamaguchi2012,yamaguchi2013chucky,Livshits2005,Yamaguchi2014,Hackett2006,Vanegue2013,Chandramohan2016BCC}). 
These approaches require the existence of known vulnerabilities or security knowledge as the guideline to formulate patterns.~As~a result, they can only identify similar but not new vulnerable code. 
Differently, we do not require any pattern inputs or prior knowledge of vulnerabilities, and can find new types of vulnerabilities.


Besides, several attempts have been made to automatically infer vulnerability patterns (e.g.,~\cite{Tan2008,Yamaguchi2015,Medeiros2016}). 
While promising, these approaches only support specific types of vulnerabilities, e.g., missing-checking vulnerabilities for~\cite{Tan2008} and taint-style vulnerabilities for~\cite{Yamaguchi2015,Medeiros2016}. However, our approach can find new types of vulnerabilities.

\noindent\textit{\textbf{Metric-Based Approaches.}} Inspired by bug prediction~\cite{Gyimothy2005, Catal2009, Hall2012, Radjenovic2013, Malhotra2015}, a number of advances have been made in applying machine learning~to~predict vulnerable code mostly at the granularity level of a source file.~In~particular, 
researchers started by leveraging complexity metrics~\cite{Shin2008a, Shin2008b, Chowdhury2011, Moshtari2013, Moshtari2016} to~predict vulnerable files.~Then, they attempted to combine complexity metrics~with more metrics such as code churn metrics and token frequency metrics~\cite{Gegick2008, Shin2011a, Shin2013, Scandariato2014, Walden2014, Zhang2015, Hovsepyan2016, Neuhaus2007, Nguyen2010, Zimmermann2010, Morrison2015,Shin2011a, Shin2013, Shin2011b}. Then, advances have been made to use unsupervised machine learning to predict bugs~\cite{liu2017code, yan2017file, yang2016effort, huang2017supervised, yan2017automated, zhang2016cross, nam2015clami, fu2017revisiting, zhou2018far} using the similar set of complexity metrics. These approaches use the similar metrics as those in bug prediction, but do not capture the~difference between vulnerable code and buggy code, which hinders the effectiveness. Moreover, the imbalance between vulnerable and non-vulnerable code is severe, which hinders the applicability of machine learning to vulnerable code identification. Instead, our approach specifically derives a set of vulnerability metrics to help identify vulnerable functions.



\noindent\textit{\textbf{Vulnerability-Specific Static Analysis.}} Researchers have attempted to detect specific~types of vulnerabilities via static analysis; e.g., buffer overflows~\cite{Evans2002,Zitser2004}, format~string~vulnerabilities~\cite{Shankar2001,Evans2002}, SQL injections~\cite{Xie2006,Jovanovic2006,Dahse2014}, cross-site scripting \cite{Jovanovic2006,Dahse2014,Lekies2013} and client-side validation vulnerabilities~\cite{Saxena2010}. While~they are effective at detecting~specific types~of vulnerabilities, they often fail to be applicable to a wider range~of vulnerability types. Moreover, they often require heavyweight program analysis techniques. 
Differently, our approach is designed to be generic and lightweight.

%% file: sec07-conclusions.tex

\section{Conclusions}\label{sec:conclusions}

We have proposed and implemented a generic, lightweight and extensible framework, named \tool, to identify potential vulnerable code~at~the function level through two sets of systematically derived program metrics. Experimental results on 11 real-world projects have demonstrated the effectiveness, scalability and applications of \tool.